\documentclass[traditabstract,a4paper]{aa}
\usepackage{txfonts} 
\usepackage{natbib}
\bibpunct{(}{)}{;}{a}{}{,} 
\usepackage{graphicx}
\usepackage{fixltx2e}
\usepackage{multirow}
\usepackage{overpic}
\usepackage{array}
\usepackage[english]{babel} 
\usepackage[utf8]{inputenc} 
\usepackage[T1]{fontenc}
\usepackage[mathscr]{eucal}
\usepackage{epstopdf}
\usepackage{epsf,color}
\usepackage{newlfont}
\usepackage{textcomp}
\usepackage{times}
\usepackage{mathrsfs}
\usepackage{xcolor}

\begin{document}

\title{A wide field-of-view low-resolution spectrometer at APEX: instrument design and science forecast}

\author{
The CONCERTO collaboration:
P.~Ade \inst{1},
M.~Aravena\inst{2},
E.~Barria \inst{3,4},
A.~Beelen \inst{5},
A.~Benoit \inst{3,4},
M.~B\'ethermin \inst{5},
J.~Bounmy \inst{6,4},  
O.~Bourrion \inst{6,4},  
G.~Bres \inst{3,4},
C.~De Breuck \inst{7},
M.~Calvo \inst{3,4},
Y.~Cao \inst{5},
A.~Catalano \inst{6,4},
F.-X.~D\'esert \inst{8,4},
C.A Dur\'an \inst{9},
A.~Fasano \inst{3,4},  
T.~Fenouillet \inst{5},
J.~Garcia  \inst{5}, 
G.~Garde \inst{3,4}, 
J.~Goupy \inst{3,4},
C.~Groppi \inst{10},
C.~Hoarau \inst{6,4},  
G.~Lagache \inst{5},
J.-C.~Lambert \inst{5},
J.-P.~Leggeri \inst{3,4},
F.~Levy-Bertrand \inst{3,4},
J.~Macias-Perez \inst{6,4},
H.~Mani\inst{10},
J.~Marpaud \inst{6,4},
P.~Mauskopf \inst{10},
A.~Monfardini \inst{3,4},
G.~Pisano \inst{1},
N.~Ponthieu \inst{8,4}, 
L.~Prieur \inst{5},
S.~Roni \inst{6},
S.~Roudier \inst{6},
D.~Tourres \inst{6,4},
C.~Tucker \inst{1} 
}

\offprints{monfardini@neel.cnrs.fr, guilaine.lagache@lam.fr}

\institute{
Astronomy Instrumentation Group, University of Cardiff, The Parade, CF24 3AA, United Kindgom
\and
N\'ucleo de Astronom\'ia, Facultad de Ingenier\'ia y Ciencias, Universidad Diego Portales, Av.  Ej\'ercito 441, Santiago, Chile
\and
Univ. Grenoble Alpes, CNRS, Grenoble INP, Institut N\'eel, 38000 Grenoble, France
\and 
Groupement d'Interet Scientifique KID, 38000 Grenoble and 38400 Saint Martin d'H\'eres, France
\and 
Aix Marseille Universit\'e, CNRS, LAM (Laboratoire d'Astrophysique de Marseille), F-13388 Marseille, France
\and 
Univ. Grenoble Alpes, CNRS, LPSC/IN2P3, 38000 Grenoble, France
\and 
European Southern Observatory, Karl Schwarzschild Straße 2, 85748 Garching, Germany
\and
Univ. Grenoble Alpes, CNRS, IPAG, 38400 Saint Martin d'H\'eres, France
\and
European Southern Observatory, Alonso de Cordova 3107, Vitacura, Santiago, Chile
\and
School of Earth and Space Exploration and Department of Physics, Arizona State University, Tempe, AZ 85287, USA
}  
 
 \abstract
 {
 \emph{Context.} 
 Characterise the large-scale structure in the Universe from present times to the high redshift epoch of reionisation is essential to constraining the cosmology, the history of star formation and reionisation, measuring the gas content of the Universe and obtaining a better understanding of the physical process that drive galaxy formation and evolution. 
 Using the integrated emission from unresolved galaxies or gas clouds, line intensity mapping (LIM) provides a new observational window to measure the larger properties of structure. This very promising technique motivates the community to plan for LIM experiments.
 
\emph{Aims.} We describe the development of a large field-of-view instrument, named CONCERTO (for CarbON CII line in post-rEionisation and ReionisaTiOn epoch), operating in the range 130-310\,GHz from the APEX 12-meters telescope (5100\,meters above sea level). CONCERTO is a low-resolution spectrometer based on the Lumped Element Kinetic Inductance Detectors (LEKID) technology. Spectra are obtained using a fast Fourier Transform Spectrometer (FTS), coupled to a dilution cryostat with base temperature of 0.1\,K. Two 2 kilo-pixels arrays of LEKID are mounted inside the cryostat that also contains the cold optics and the front-end electronics. 

\emph{Methods.} We present in detail the technological choices leading to the instrumental concept, together with the design and fabrication of the instrument and preliminary laboratory tests on the detectors. We also give our best estimates of CONCERTO sensitivity and give predictions for two of the main scientific goals of CONCERTO, i.e. a [CII]-intensity mapping survey and observations of galaxy clusters.

\emph{Results.} We provide a detail description of the instrument design. Based on realistic comparisons with existing instruments developed by our group (NIKA, NIKA2, and KISS), and on laboratory detectors characterisation, we provide an estimate of CONCERTO sensitivity on sky. Finally, we describe in detail two out of the main science goals offered by CONCERTO at APEX. 

}
 
\authorrunning{The CONCERTO collaboration}
\titlerunning{A wide field-of-view low-resolution spectrometer at APEX}

\maketitle

\section{Introduction} \label{sec1}

Modern imaging and polarimetry cameras, at  millimetre and sub-millimetre wavelengths, are currently operating on large (e.g. D $>$ 10~meters) single-dish telescopes. The main goal of such instruments is to map, at relatively high angular resolution (e.g. 5-30 arcseconds) large portions of the sky (e.g. several deg$^2$) with high sensitivity (e.g. RMS$_{\mathrm{MAP}} \lesssim$ 1\,mJy). Polarised emissions are also measured with similar specifications. 

In this context, the dual-band NIKA2 camera represents the first kilo-pixels instrument operating at these wavelengths based on the Kinetic Inductance Detectors (KID) technology \citep{nika2_1,nika2_2}. The particular flavour of KID used for NIKA2 (and CONCERTO) are front-illuminated Lumped Elements KID (LEKID) \citep{Doyle2010}. They consist in inductor-capacitor (LC) superconducting planar resonators made by a long meandered inductor (wire) terminated at both ends by an inter-digitated capacitor. NIKA2 supersedes previous cameras based on Transition Edge Sensors (referred to as TES bolometers) in the frequency range 150-360\,GHz, such as MAMBO2 at IRAM \citep{kreysa1998}, LABOCA at APEX \citep{Siringo2009} and SCUBA-2 at the JCMT \citep{holland2013}.

In order to extend the capabilities of the existing instruments, and open new observational windows of the millimetre sky, the spectral dimension has to be added, without sacrificing the instantaneous field-of-view. The large field-of-view and the mapping speed are, actually, the main asset of single dish telescopes when compared to variable baseline interferometers like ALMA\footnote{https://almascience.eso.org} or NOEMA \citep{NOEMA2020}. For this reason, we are developing a millimetre-wave low spectral resolution (R=$\nu / \Delta \nu \leq 300$) spectrometer with an instantaneous field-of-view of 20\,arcminutes. In order to preserve the angular resolution, at frequencies of around 300\,GHz and assuming a 10-meters class telescope, a focal-plane containing around 2,000 spatial pixels is needed. To achieve these figures, we adopt a room-temperature Martin-Puplett Interferometer (MpI) \citep{MPI70} coupled to a large field millimetre-wave camera. The instrument, named CONCERTO, has been designed to interface with the Atacama Pathfinder EXperiment (APEX) 12-meters telescope \citep{Gusten2006}. A pathfinder instrument, named KISS and based on the same concept, has been built by our collaboration and deployed in November, 2018, at the Teide Observatory \citep{KISS2020}. 

One important science driver that has motivated our developments is the study of [CII] emission line at high redshift. [CII] is among the brightest lines originating from star-forming galaxies and a reliable tracer of star formation on global scales.
With CONCERTO at APEX, we will map in three dimensions the fluctuations of the [CII] line intensity in the reionisation and post-reionisation epoch ($z\gtrsim5$). This technique, known as "intensity mapping", will allow to answer the questions of whether dusty star-formation contributes to early galaxy evolution, and whether [CII]-emitters play an important role in shaping cosmic reionisation. The dedicated [CII] survey will provide a (spatial-spectral) data cube in which intensity is mapped as a function of sky position and frequency. The 3-D fluctuations are then studied in Fourier space with the power spectrum. 
The [CII] survey will also be sensitive to the CO intensity fluctuations arising from $0.3<z<2$ galaxies, giving the spatial distribution and abundance of molecular gas over a broad range of cosmic time.
[CII] intensity mapping is also one of the main goals of CCAT-prime \citep{choi2020} and TIME \citep{crites2014}, two experiments based on different technologies than CONCERTO: gratings and TES (Transition Edge Sensors) bolometers for TIME, KID and Fabry-Perot interferometers for CCAT-p. 
In addition to the main [CII] survey, we expect CONCERTO to bring a significant contribution in a number of areas, including the study of galaxy clusters (via the thermal and kinetic SZ effect), the observation of local and intermediate-redshift galaxies, and the study of Galactic star-forming clouds.  In this paper, we detail the main goals of the [CII] intensity mapping and galaxy clusters surveys. 

The paper is organised as follow. We present the instrumental concept, design and preliminary results in Sect.\,\ref{sec2}. The discussion includes KID detectors, cryogenics and optics. In Sect.\,\ref{sec3}, we describe the first laboratory tests (still on-going for a full characterisation). In Sect.\,\ref{sec4}, we present the sensitivity estimates, while Sect.\,\ref{sec5} is dedicated to the [CII] intensity mapping and SZ surveys.

\section{CONCERTO instrument}\label{sec2}

\begin{figure}[ht]
\begin{center}
\includegraphics[width=9cm, angle=0]{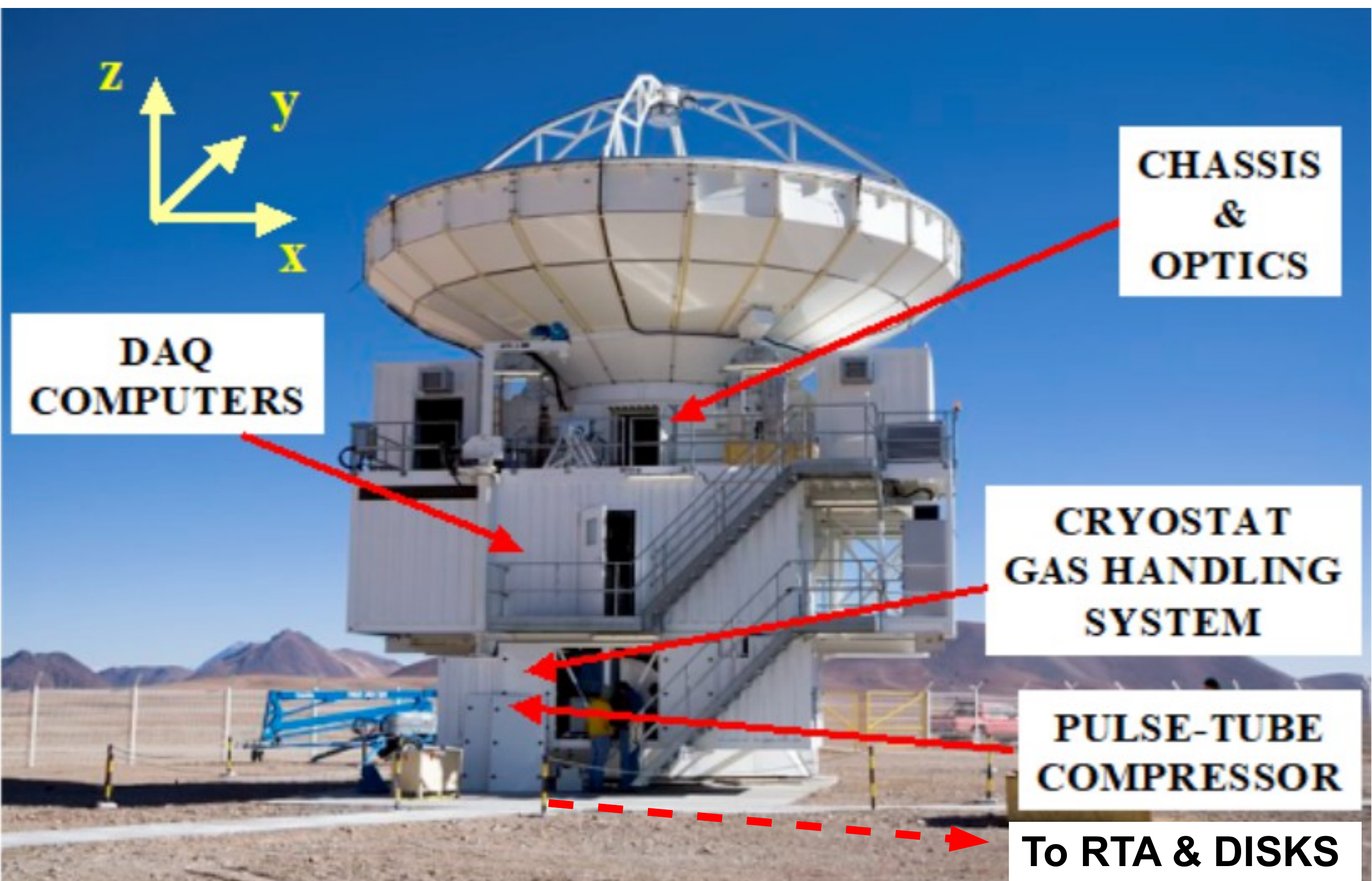}
\end{center}
  \caption{Location of the CONCERTO sub-systems: the chassis and optics in the C-cabin, the Data AcQuisition computers (DAQ) in the middle container ("instrumentation container") and  the gas handling system and the pulse-tube compressor in the bottom container ("compressors room"). Real Time Analysis (RTA) computers and hard disks are not in the telescope tower.}
\label{whole}
\end{figure}

CONCERTO has been specifically designed to fit into the Cassegrain cabin (C-cabin) of the APEX telescope. It is composed of two main components: the so-called "chassis" and the "optics box". The "chassis" includes the camera (cryostat), the MpI interferometer, the readout and control electronics. The "optics box" includes a number of mirrors and polarisers and a cold reference for the MpI. 
In Table \ref{tab1} we summarise the main instrument characteristics. The location of CONCERTO sub-systems are shown in Fig.\,\ref{whole}.

\begin{table}[ht]
{
\begin{center}
\begin{tabular}{cc}
\hline \hline
Telescope primary mirror diameter [m] & 12 \\
Field-of-view diameter [arcmin] & 20 \\
Absolute spectral resolution [GHz] & $\geq$ 1 \\
Relative spectral resolution R [\#] & 1-300 \\
Frequency range HF $\mid$ LF [GHz] & 195-310 $\mid$ 130-270 \\
Pixels on Sky HF $\mid$ LF [\#] & 2152 $\mid$ 2152 \\
Angular resolution HF $\mid$ LF [arcsec] & 20-32 $\mid$ 23-45 \\
Average angular resolution HF $\mid$ LF [arcsec] & 26 $\mid$ 34 \\
Instrument geometrical throughput [sr\,m$^2$] & 2.5$\times$10$^{-3}$\\
Single Pixel geometrical throughput [sr$\,$m$^2$] & 1.16$\times$10$^{-6}$\\
Data rate [MBytes/sec] & 128 \\
\hline \hline
\end{tabular}
\end{center}
}
\caption{Main characteristics of CONCERTO. For comparison, the NIKA2 instrument (6.5' field-of-view, 30-meters telescope) geometrical throughput (A$\times\Omega$), characterising the camera collecting power, is 1.7$\times$10$^{-3}$~sr$\,$m$^2$. Concerning the overall optical transmission of CONCERTO compared to NIKA2, we refer the reader to Sect.\,\ref{sec4.1}.} 
\label{tab1}
\end{table}

In this section, we describe in detail the camera and its content (Sect.\,\ref{subsec2-1}), the MpI (Sect.\,\ref{subsec2-2}), and the chassis and related electronics (Sect.\,\ref{subsec2-3}). The room temperature optics, including the cold reference source, is presented in Sect.\,\ref{subsec2-4}. Section\,\ref{subsec2-5} is devoted to a brief description of the CONCERTO hardware components located elsewhere than in the telescope tower. The installation at the telescope is the purpose of Sect.\,\ref{subsec2-6}.

\subsection{The camera}\label{subsec2-1}

CONCERTO camera is based on a cryogenic-liquid-free custom dilution cryostat. The dilution insert and the pulse-tube orientation, in particular, have been specifically designed to allow the rotation of the cryostat axis following telescope movements. The cryostat is optimised for the range of telescope elevations (EL) comprised between 30 and 90~degrees. The best working point is achieved for EL=60~degrees.

\begin{figure}[ht]
\begin{center}
\includegraphics[width=9cm, angle=0]{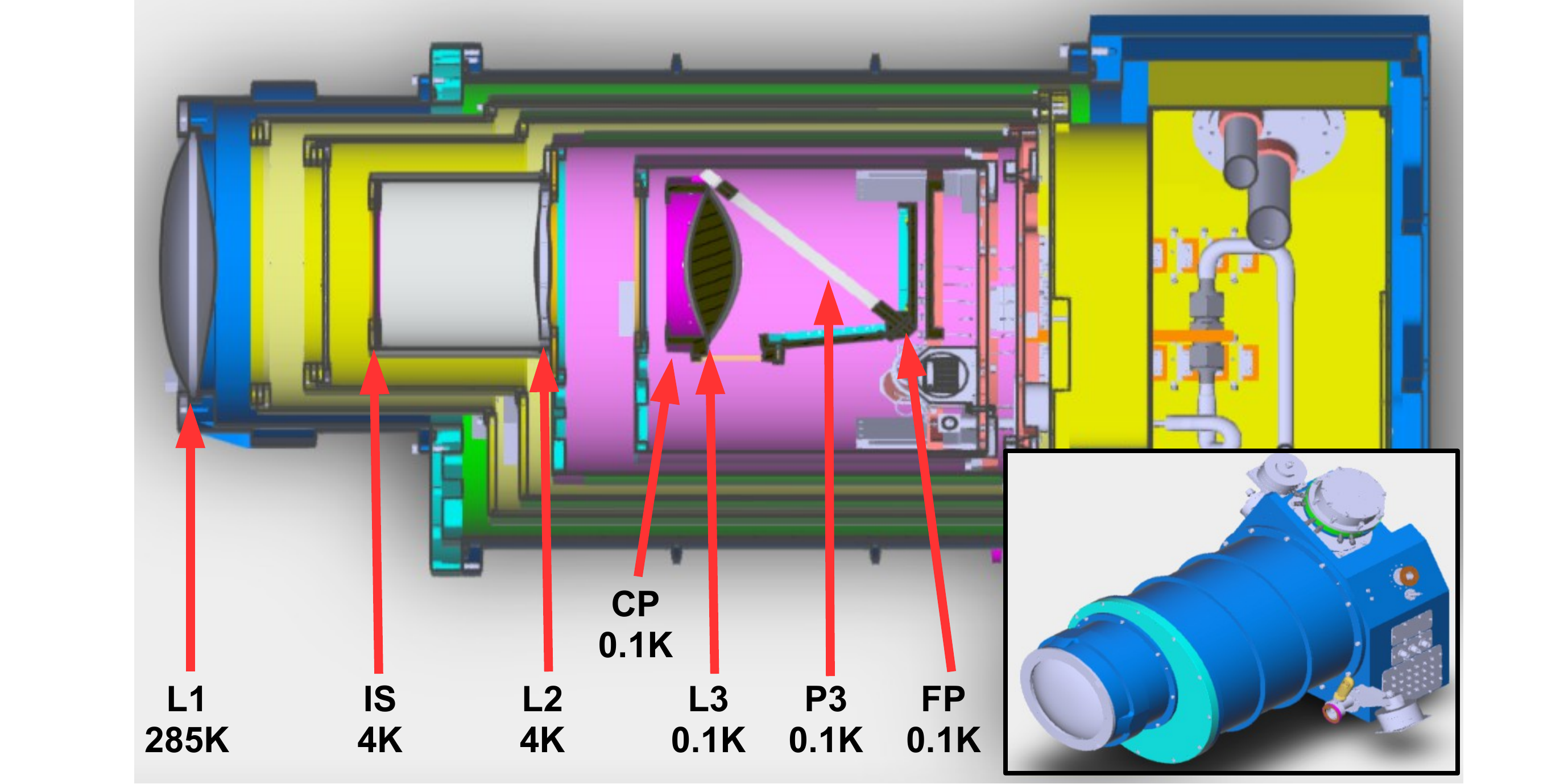}
\end{center}
  \caption{CONCERTO camera cross-section and 3-D view (inset). The positions of the three HDPE lenses are shown (L1, L2, L3), together with the Image Stop (IS), the Cold Pupil (CP), the cold polariser (P3) and the two focal planes (FP). The diameter of L1 is around 250~mm.}
\label{cryostat}
\end{figure}

The main camera optical features, shown in Fig. \ref{cryostat}, are the Image Stop (IS) at a temperature of 4\,K and the Cold Pupil (CP) at the base temperature of around 0.1\,K. Three HDPE (High Density PolyEthilene) lenses are used in the camera: L1 (room temperature), L2 (4\,K) and L3 (0.1\,K). In order to analyse the polarised signal, the last polariser of the MpI (P3) is placed just in front of the LEKID arrays (FP) at base temperature. P3 is a custom wire-grid polariser. It is realised on a 12~$\mu$m-thick Polyimide membrane, and with Copper wires with a pitch of 50\,$\mu$m. A number of IR-blocking (thermal) and metallic multi-meshes filters \citep{Pisano} are mounted at different stages. In particular, we have thermal filters on the warmest stages (room temperature to 50\,K), low-pass multi-mesh filters at the intermediate temperatures (50\,K to 1\,K) and band-defining filters at base temperature, i.e. just in front of each focal-plane array. A specially blackened baffle is installed at 4\,K, between IS and L2, in order to suppress the stray-light. 

Since LEKID are sensitive to variations of magnetic fields, they have to be protected by a multi-stage B field screen. Four concentric high-permittivity alloys (i.e. mu-metal and cryogenics variations) cylinders are installed at 300\,K, 50\,K (double screen) and 4\,K. An additional superconducting screen will be wrapped around the focal planes section.

The focal plane arrays are microstrip-coupled LEKID similar to those used for NIKA2 \citep{nika2_1}.  Six excitation/readout lines (feed-lines) are needed to readout each of the 2152 pixels array. A total of twelve pairs of coaxial cables are thus running into the cryostat. The pixels design itself is derived from NIKA \citep{nika1}. The LEKID details have been optimised to meet the CONCERTO specifications. In particular, the shape of the meander and its coupling quality factor have been adjusted to the target range of frequencies and expected background. The coupling quality factor is designed to be Q$_c\approx\Delta f_{-3dB} / f_0 \approx$ 2.5$\times$10$^4$, with $\Delta f_{-3dB}$ the typical width of the resonance under dark conditions. The thickness of the dielectric substrate is calibrated in order to maximise the quantum efficiency, and is in the range 100-120\,$\mu$m. 

\begin{figure}[ht]
\begin{center}
\includegraphics[width=7.5cm, angle=0]{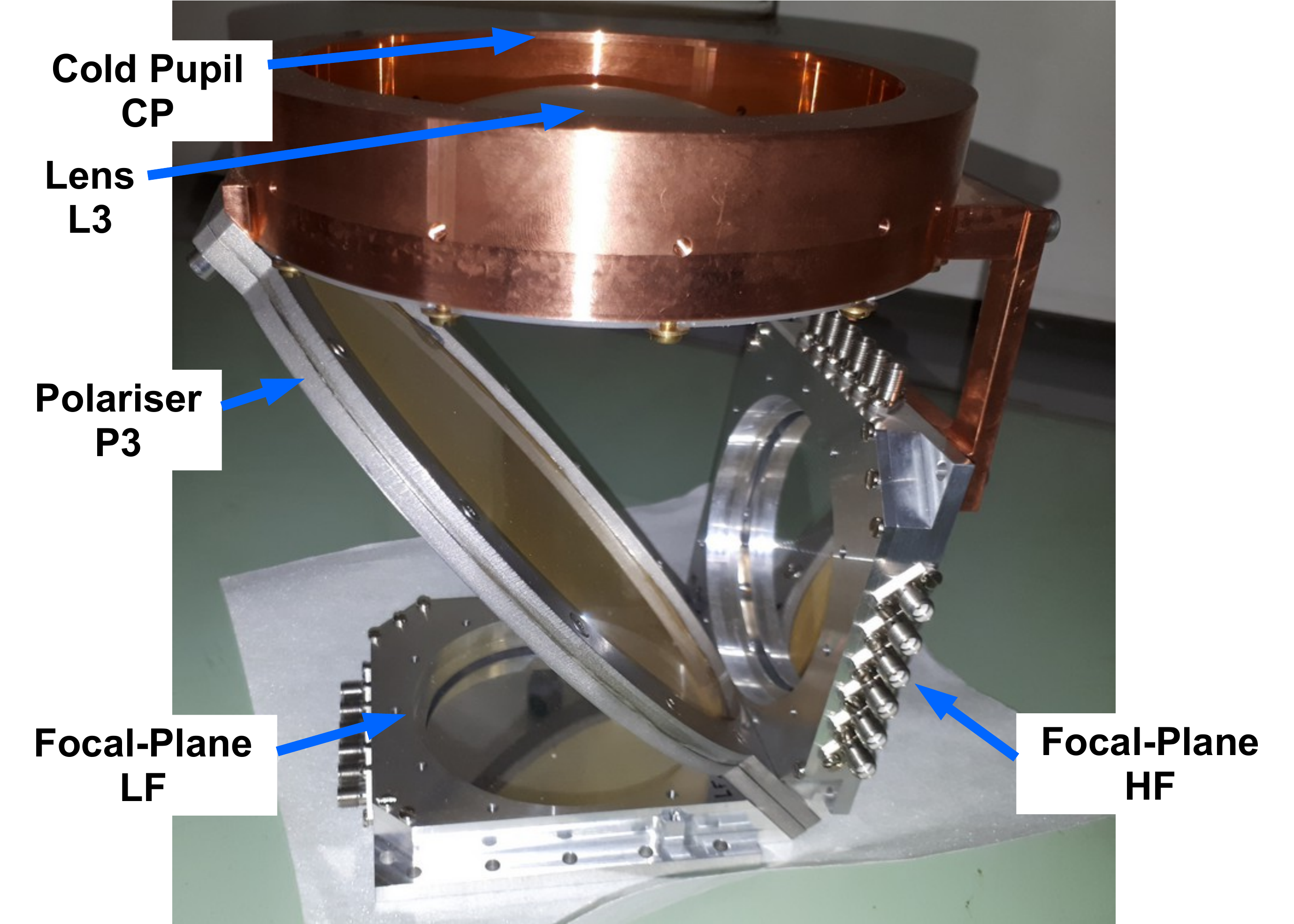}
\end{center}
 \caption{Picture of the first "100\,mK block" including the cold pupil, the L3 lens, the P3 polariser and the two arrays (HF and LF) containing 2152~pixels each. The arrays holders are in this case realised in Aluminium. A version of the block with Copper holders is also available.}
\label{array}
\end{figure}

The fabrication process \citep{dicing2016} looks straightforward when compared to competing detectors having similar performance. The substrate, a 100-millimetres high-purity mono-crystalline Silicon wafer\footnote{https://www.sil-tronix-st.com/en/}, is prepared in the deposition chamber by a soft ion milling. The superconducting film deposition, Aluminium with thickness of 20\,nm, is achieved by e-beam evaporation and under a residual chamber pressure of 5$\times$10$^{-8}$~mbars. The deposition rates is fixed at 0.1\,nm/sec. The UV photo-lithography step is based on a positive resist, and is followed by wet etching. The etching is done using a standard Aluminium etching solution based on phosphoric acid. The diced detectors arrays are packaged in custom holders and bonded, via 17\,$\mu$m Aluminium wires, to the 50-Ohms micro-strip launchers. Those are then tin-soldered to the inner pin of the SMA\footnote{SubMiniature version A} feed-throughs. 

The front-end electronics stage is installed in the cryostat, at a temperature of 4\,K. It is made of a series of twelve low-noise amplifiers (LNA) operating at the resonance frequencies\footnote{http://thz.asu.edu/products.html}, i.e. in the range 1.5-2.5\,GHz. A second stage of cryogenic amplification, i.e. twelve commercial LNA, has been added on the 50\,K cryogenic stage to simplify the room-temperature electronics and reduce its power consumption. The connections between the cold electronics stages, the arrays and the SMA vacuum feed-throughs plate (see Fig.\,\ref{cryostat}), are ensured by commercial semi-rigid cryogenic coaxial cables. In particular, we adopt NbTi superconducting coaxial cables for the portion connecting the output of the LEKID arrays to the input of the front-end amplifiers. Fixed attenuators are mounted on each input line, at the 4\,K stage. The overall electrical gain of each radio-frequency line to/from the room temperature electronics has been measured and is about +25\,dB.

\subsection{The Martin-Puplett Interferometer}\label{subsec2-2}

\begin{figure}[ht]
\begin{center}
\includegraphics[width=9cm,angle=0]{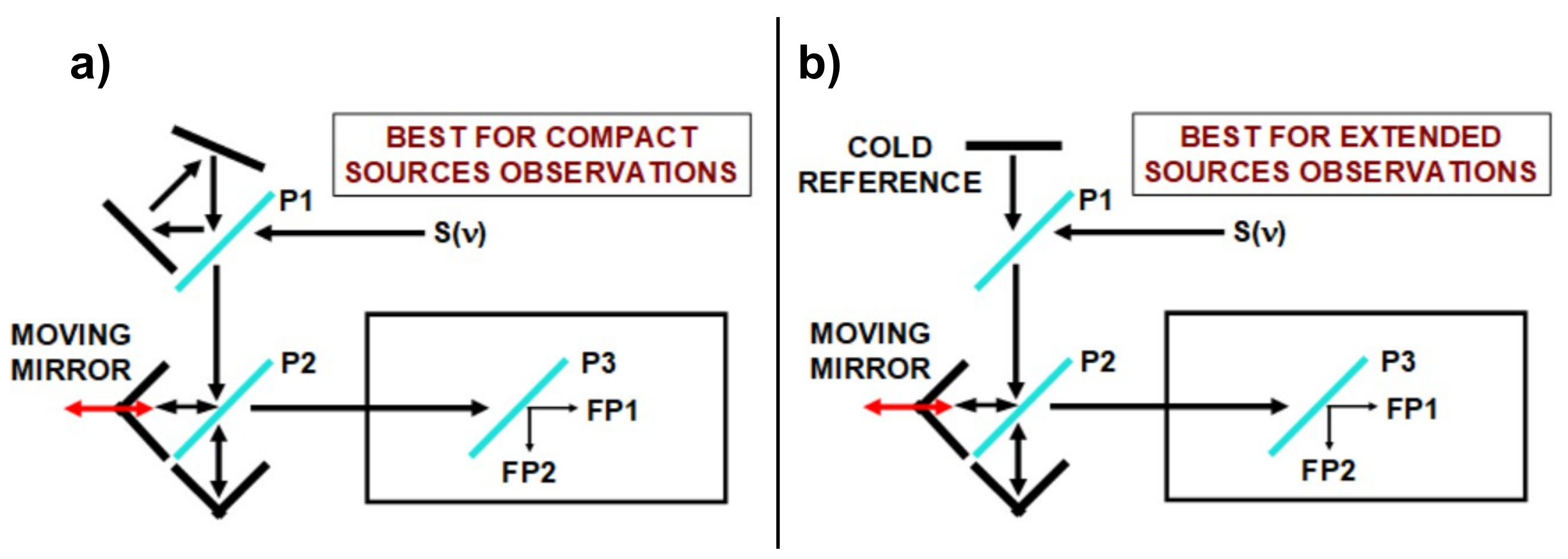}
\end{center}
\caption{Schematics of the MpI concept. Two options are shown for the reference source: a) a de-focused image of the instantaneous field-of-view; b) a cold reference. The polariser P1 provides the needed polarised input to the MpI. P2 is the beam splitter defining the two arms while P3, in the cryostat, dispatches the two projections of the polarised signal to the focal-plane arrays. The incoming beam is represented by a spectral distribution S($\nu$).}
\label{MPIreference}
\end{figure}

The MpI is a particular kind of Fourier Transform Spectrometer (FTS). It is capable of measuring the differential spectrum of a source, with respect to a given reference. The key elements are three polarisers (P1-beam divider, P2-splitter and P3-analyser), two (fixed/variable) arms, and two (fixed/moving) rooftop mirrors (see Fig.\,\ref{MPIreference}). This technique is widely adopted in the millimetre and sub-millimetre domains, mostly for laboratory characterisations, but also, in the past, for narrow\footnote{In this context, the field-of-view has to be expressed in number of beams. The number of beams is around 2,000 for CONCERTO and 300 for KISS, to be compared to few tens at maximum for the previous instruments.} field-of-view observations from space \citep{firas99, Griffin2010}. The first examples of wide field instruments making use of an MpI to obtain spectral information are the stratospheric balloon OLIMPO \citep{OLIMPO2014} and the already mentioned KISS ground-based spectro-photometre \citep{Fasano2020}. 

To achieve a spectral resolution better than 1\,GHz, the maximum range $\Delta$l$_{max}$ of the moving rooftop mirror has to larger than 75\,mm. In CONCERTO, the motors can move by up to 90\,mm.
The range $\Delta$l spanned by the interferogram can be adjusted, on a scan-by-scan basis and depending on the science target, from zero to the maximum. The spectral resolution will then be 
$$\Delta \nu = \frac{c}{4 \times \Delta l}\,.$$
The optical path difference is thus OPD=$2\Delta l$. Leaving the rooftop mirror stopped at the zero-path difference position results in using CONCERTO as a broad- and dual-band large field-of-view imager. 

\begin{figure}[ht]
\begin{center}
\includegraphics[width=9cm, angle=0]{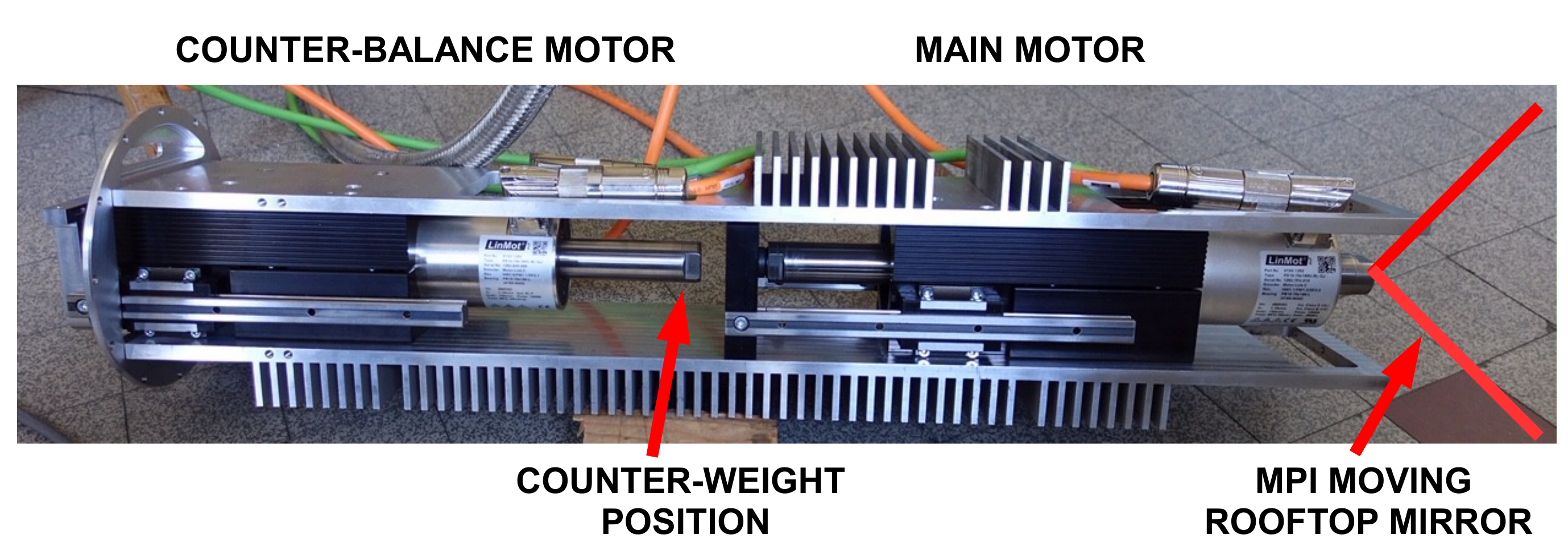}
\end{center}
\caption{CONCERTO double-motors MpI. Two identical linear motors, developing a force of $\geq$1000\,N each, are acted on in counter-phase to null the total momentum.}
\label{MPI3D}
\end{figure}

The distinctive feature of the CONCERTO (and KISS) MpI is the combination of the speed of movement of the rooftop mirror and its size/mass. In order to avoid atmospheric drifts during a single interferogram, the mechanical frequency of the motors is set to around 4\,Hz, i.e. 8 full interferograms (and spectra) per second are produced by each of the pixels. The lateral size of the mirror to be moved exceeds 0.5\,meters, for a mass exceeding 3\,kg. In order to counter-balance the linear momentum associated to such a moving mass, a second motor, with an equivalent mass, oscillates with an opposite instantaneous velocity. With a maximum force around 1000\,N, the acceleration that can be imposed to the moving mass, including the motor piston, exceeds 100\,m/s$^2$. The theoretical curve that is commanded to the motor is a square wave, i.e. constant speed for both ways and maximum acceleration at the turn-backs. The real curve is of course smoothed out by the finite acceleration near the extremes. We present a picture of the system that has been built for CONCERTO in Fig.\,\ref{MPI3D}.

\subsection{The "chassis" and electronics}\label{subsec2-3}

The chassis is a single, compact support structure to which many of the core components of CONCERTO are attached. These include the camera itself, the MpI motors and moving mirror, and the electronics boards, along with a large number of modules devoted to the monitoring and control of the instrument. The chassis has been designed and fabricated to match the constraints related to the limited space available in the C-cabin. It allows installing multiple sub-systems of CONCERTO as a single element inside the APEX C-cabin. The chassis is laboratory pre-mounted and can slide through the C-cabin door.

Five microTCA\footnote{https://en.wikipedia.org/wiki/MicroTCA} racks mounted on the side of the chassis host the 12 Advanced Mezzanine Cards (AMC) used to readout the two arrays. The cards have an architecture similar to those used for NIKA2 \citep{bourrion2012}, but they have been improved to be able to generate up to 400 excitation tones spanning 1\,GHz bandwidth. The data acquisition rate has been increased from less than 100\,Hz up to 4\,kHz, in order to properly sample the interferograms generated by the MpI. 
The calibration strategy has been inherited from the NIKA and NIKA2 instruments \citep{calvo}, but the continuous frequency modulation used there is no longer viable because of the high-sampling rate. As a consequence, in CONCERTO, the position and shape of each resonance circle, which is used to calibrate the data, is reconstructed by sampling three points around the resonance: $f_0$, $f_0+\delta f$, $f_0-\delta f$. The $\delta f$ is much smaller than the resonance width, and of the order of a few kHz. The calibration step is performed at the beginning of each interferogram, while the MpI rooftop mirror is changing the direction of its motion \citep{KISS2020}. This approach represents in our opinion the best trade-off between optimal calibration and observing efficiency, i.e. the fraction of time that is devoted to science data stream.

\begin{figure}[ht]
\begin{center}
\includegraphics[width=9cm, angle=0]{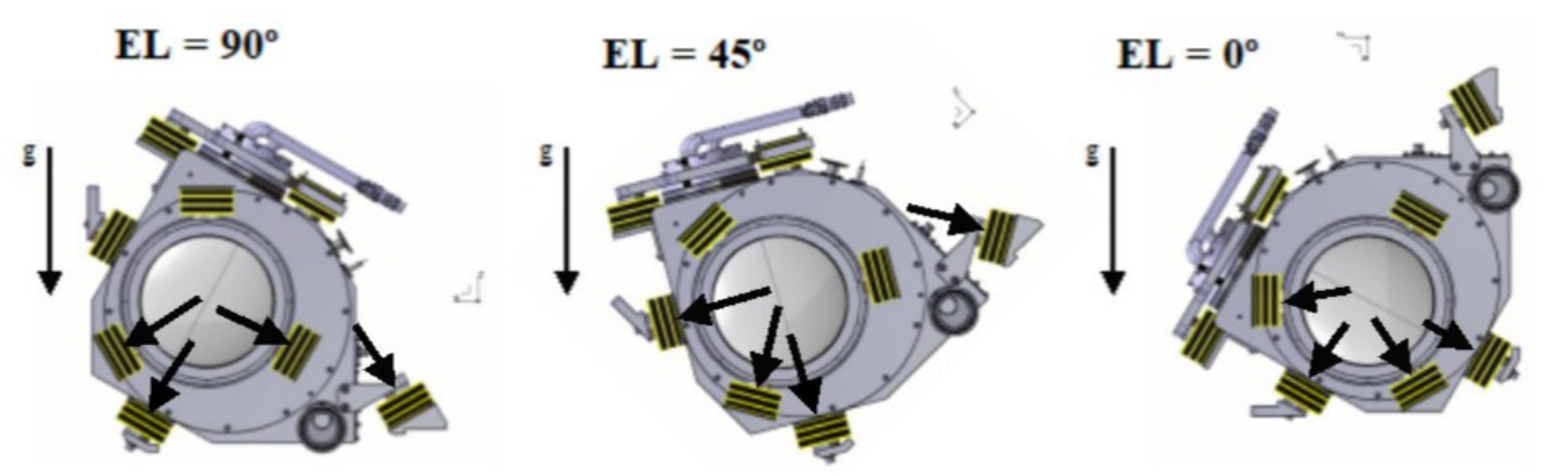}
\end{center}
\caption{Rotation of the camera (and chassis) following the telescope elevation (EL). The position of the eight soft rubber pneumatic actuators is shown. Two of them are dedicated to the pulse-tube head. For the six remaining, we indicate (with black arrows) those at action for three representative elevation cases.}
\label{rotation}
\end{figure}

The moving elements inside the chassis, in particular the MpI motors and the gas flowing in the Pulse Tube head, generate vibrations which could affect the detector performances. In order to suppress their propagation to the focal plane, the camera is fixed to the chassis only via a series of soft rubber pneumatic actuators, which strongly dampen the vibrations. Furthermore, the pressure inside the actuators (eight in total, with different orientations, as shown in Fig.\,\ref{rotation}) is constantly adjusted by a dedicated software, so that the cryostat position and axis, monitored by means of linear position transducers, are kept constant independently of the telescope elevation. The cryostat position adjustment, requiring a few seconds to complete, is done automatically after each re-pointing and, upon request, between two subsequent observing blocks (scans). The position is, on the other hand, monitored in real time even during scans.

\subsection{The cabin optics and the cold reference (optics box)}\label{subsec2-4}

The first CONCERTO element along the optical axis, after the telescope mirrors (M1 and M2), is the M3 foldable mirror mounted on the chassis. With a diameter of 900\,mm, it is the largest among the CONCERTO mirrors. The next mirror M4, attached to C-cabin upper ring, reflects the beam toward the C-cabin floor and directly into the so-called "optics box". A first virtual image is generated, by the combination M1-M2-M3-M4, before M5. This virtual image plane will be used for some of the CONCERTO qualification tests.

\begin{figure}[ht]
\begin{center}
\includegraphics[width=9cm, angle=0]{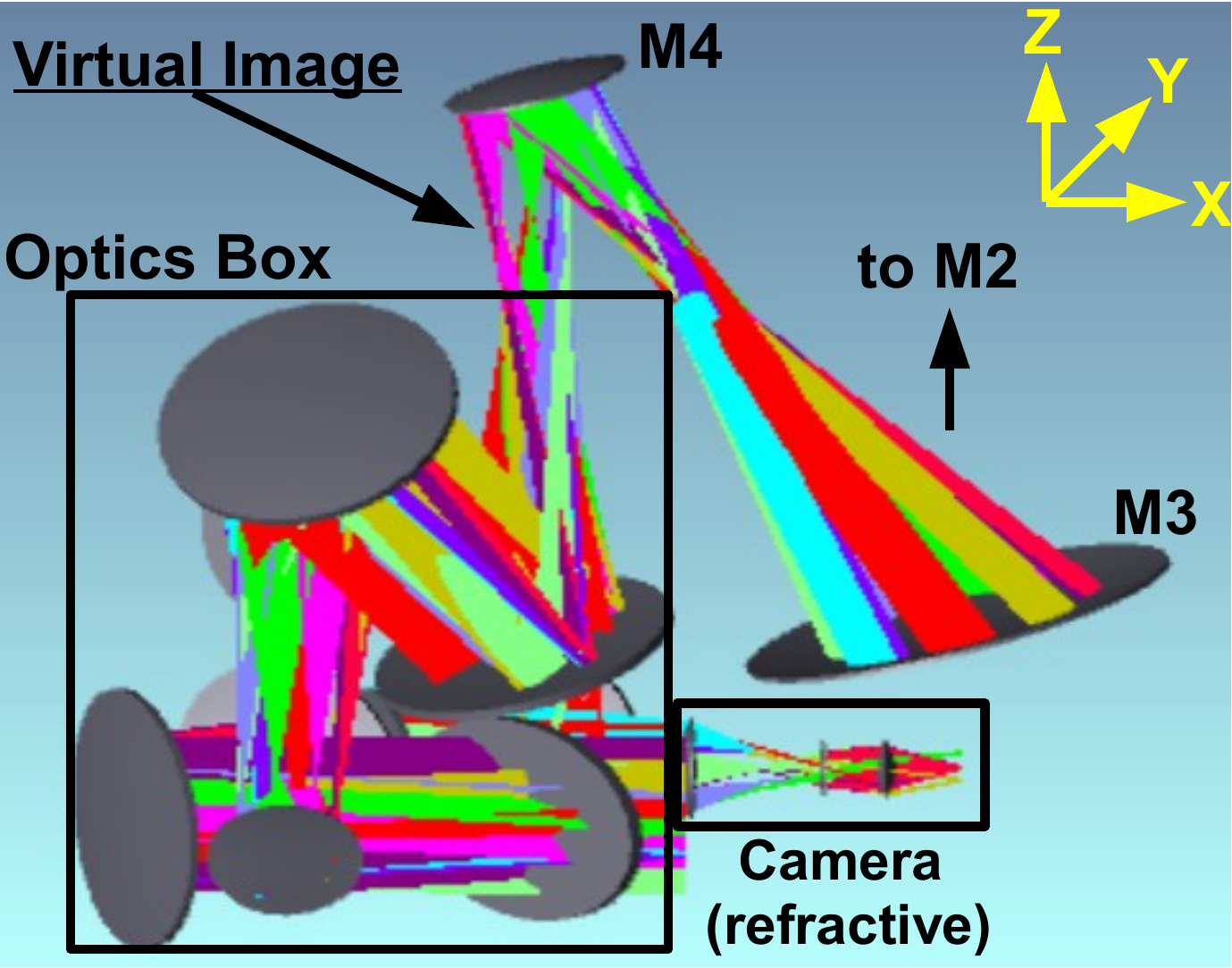}
\end{center}
\caption{3-D view of the CONCERTO optics, evidencing the M3 mirror interfaced to the APEX sub-reflector (M2). M4 is attached to the ceiling of the C-cabin and represents the only reflective optics component, with M3 and the MpI rooftop moving mirror, outside of the "optics box". Completing the optical chain: a large number of mirrors (M5 to M11), the two polarisers P1 and P2, and the cold reference optics.}
\label{optics1}
\end{figure}

The "optics box" includes a large number of mirrors (M5 to M11), the two polarisers P1 and P2 (Fig.\,\ref{MPIreference}) and the part of the optics providing the cold reference for the MpI. It also includes the fixed rooftop mirror of the interferometer. A general 3-D view is shown in Fig.\,\ref{optics1}.

The mirrors are held at the C-cabin temperature, that is regulated at 11\,$^{\mathrm o}$C=284\,K. We expect an emissivity of the order of 1\% per mirror, equivalent to an additional background of about 3\,K per surface, so not smaller than 30\,K in total.
The stability of the temperature in the cabin is $\pm$1\,K. This means that, for an emissivity of 1\%, the effective background temperature variation induced per mirror is around 10\,mK. Considering roughly 10 mirrors, this translates to a total effective background temperature variation of 0.1\,K. The KID detectors, with a NET of the order of $\approx$\,mK$\cdot\sqrt{s}$ per pixel, are sensitive to this drift that will produce a correlated signal on all the pixels. However, these instabilities are harmless since they are slower and smaller than the atmospheric fluctuations sitting on top.

\begin{figure}[ht]
\begin{center}
\includegraphics[width=9cm, angle=0]{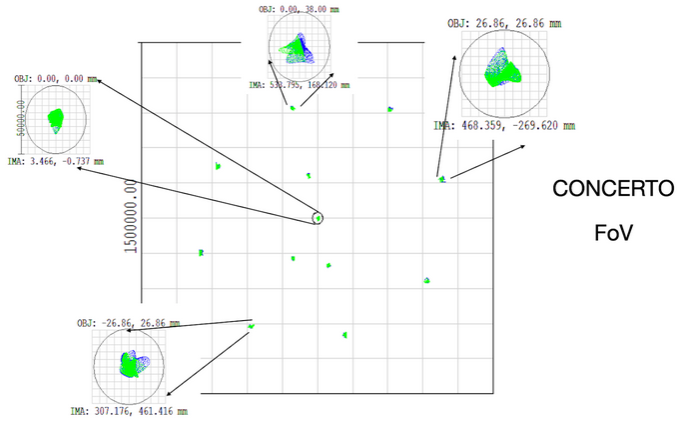}
\end{center}
\caption{Simulated focal plane image for the full, 20\,arcmin, field-of-view. Each spot is shown for the two extreme positions of the roof mirror, i.e. green (0\,mm) and blue (90\,mm). The black circles represent the Airy disks. It is clear that the "walking" of the beam is well contained in the diffraction disk.}
\label{walk}
\end{figure}

We performed a trade-off between the requirements related to the image quality (and the interferometry efficiency) and the lateral size of the MpI. In order to obtain a diffraction-limited combined beam for each position of the movable roof mirror (in the range 0--90\,mm), we imposed the criterion, for each field on the sky, of producing a quasi-parallel beam inside the MpI. According to the geometrical throughput conservation rule, the field-to-field divergence is thus fixed by the diameter of the beam. For the 20\,arcmin field-of-view, and considering a 12\,m primary mirror, we obtained an overall beam diameter of about 420\,mm inside the interferometer. A consequence of this method is that, imposing that the combined beam does not "walk" in the focal plane (see Fig.\,\ref{walk}), we must accept a jitter on the entrance pupil of the optical system, i.e. the "active" portion of the primary mirror. This is, in the end, the main reason why we have decided to under-sample the size of the illuminated primary mirror, to about 11 metres.\\

A remotely controllable three-positions mirror is inserted in the optics chain in order to select the type of reference input for the MpI. The three options are:
\begin{itemize}
\item \emph{Sky:} a de-focused image of the full 20\,arcminutes instantaneous field-of-view. The distinctive advantage is that the atmospheric common-mode spectrum is optically subtracted, providing a differential measurement of the astrophysics source spectrum with respect to the atmosphere along the line-of-sight. This means that, when targeting a field populated by weak sources, and at the first order, we will obtain a null interferogram. This configuration is ideally suited for compact object, i.e. angular extension smaller than 20\,arcminutes.
\item \emph{External cold black-body:} a highly-emissive ($\epsilon \geq $\,0.98) cold disk cooled down by an independent pulse-tube cryostat (T$_{BB}\approx$\,8\,K). This configuration is mostly adapted to extended emission observations, i.e. when the spectral and photometric gradients extend on average more than 20\,arcmin.
\item \emph{Cold cryostat:} a de-focused image of the CONCERTO cryostat cold (inner) parts. In other words CONCERTO "looks" into itself to find a cold auto-reference. Considering the number of optical elements (six mirrors and three lenses) lying between the three-positions mirror and the coldest stage of the cryostat, we expect an equivalent effective temperature of the order of 20\,K. This is in any case lower than the loading of the sky plus the whole optics train between M1 and the focal plane (which is not lower than 50\,K). This innovative configuration will be investigated as a simpler alternative to the external cold black body.
\end{itemize}

We stress the fact that the external cold black-body, or its alternative "cold cryostat", are not used as spectral calibrators. They simply represent cold, i.e. colder that the combined thermal emission of the atmosphere and the optical chain, references. In contrast to the "Sky" reference case, the interferogram is not expected to be null at first order. We expect in this case potential systematic effects to be minimised by the fact that both the reference and the dominant target (mostly the atmosphere in the common case of weak astrophysical sources) exhibit thermal black-body spectra.

The best choice between the three references will depend on the particular science target, the observing conditions and the still unknown systematic effects affecting this new kind of large field-of-view spectro-photometre. A crucial phase of the on-sky commissioning will be dedicated to investigating this item. We will report in further publications the results of this study, as well as a more detailed description of the CONCERTO MpI spectral reference system that is beyond the scope of the present paper.

\subsection{CONCERTO hardware outside the C-cabin}\label{subsec2-5}

On top of the elements described above and located in the C-cabin, CONCERTO is made also by modules elsewhere in the telescope tower and beyond, in particular, the commercial pulse-tube compressor (Cryomech CPA289C) and, more interestingly, the dilution cryostat Gas Handling System (GHS) and the Data AcQuisition (DAQ) and Real Time Analysis (RTA) computers. 

\begin{figure}[ht]
\begin{center}
\includegraphics[width=9cm, angle=0]{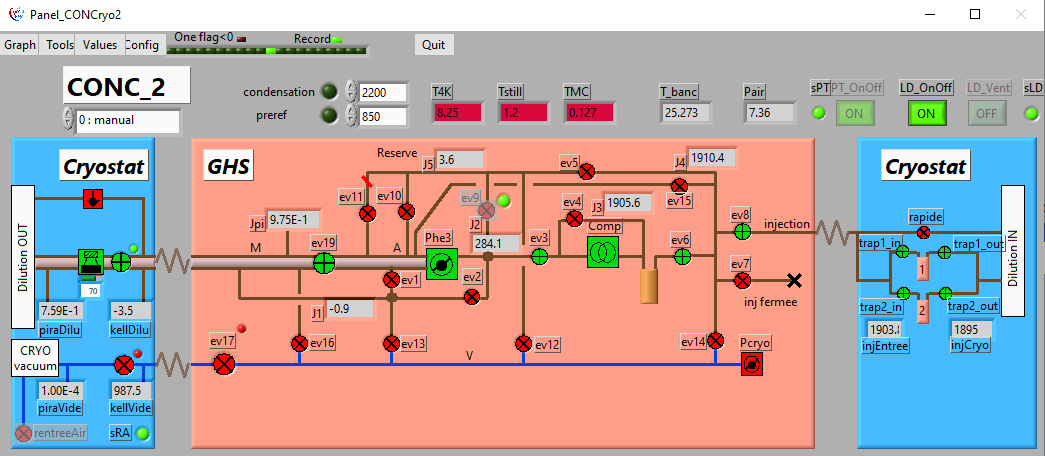}
\end{center}
\caption{The GUI used to control the dilution refrigerator circuit.}
\label{map}
\end{figure}

The GHS is composed of: a) a series of pumps and compressors for the circulation of the $^3$He-$^4$He mixture and for providing compressed air to CONCERTO, and, b) an electronic cabinet hosting a National Instrument CompactRIO\footnote{https://en.wikipedia.org/wiki/CompactRIO} real-time controller and multiple analogue and digital input-output modules. A dedicated Labview-based software is loaded on the CompactRIO. The software continuously monitors the state of the cryostat and controls all the pumps, compressors, valves and actuators of the dilution circuit. It can perform many tasks automatically, such as pre-cooling the system or putting it in a safe mode if the security thresholds are exceeded. It also acts as a server-side program for the client Graphical User Interface (GUI) that is used on remote computers. The GUI allow to easily see the state of the system and control its components (Fig.\,\ref{map}).

All the pipes and cables that are needed to interconnect the elements of CONCERTO situated in different rooms (C-Cabin, instrumentation and compressors containers) are routed through flexible hoses. The hoses protect the CONCERTO cabling from the environment and guarantee the flexibility required by the movement of the telescope. Inside the C-cabin, all the connections are centralized on a dedicated panel located on the front side of the chassis, thus easing the procedures of plugging and unplugging.

The DAQ system is located in the middle container and is connected to CONCERTO in the C-cabin, producing 128\,MBytes per second, through five dedicated Ethernet cables. The DAQ consists in two commercial computers with 48\,GB of RAM and 24 cores each. The disk storage (432\,TB) and the RTA systems are installed in the so-called "servers room", located a few tens of meters away from the telescope tower. The RTA computer has 32 cores and 512\,GB of RAM. The network connection between the DAQ and the disk/RTA is ensured by two 10 Gigabit switches and underground cables.

\subsection{Installation at the APEX telescope}\label{subsec2-6}

\begin{figure}[ht]
\begin{center}
\includegraphics[width=9cm, angle=0]{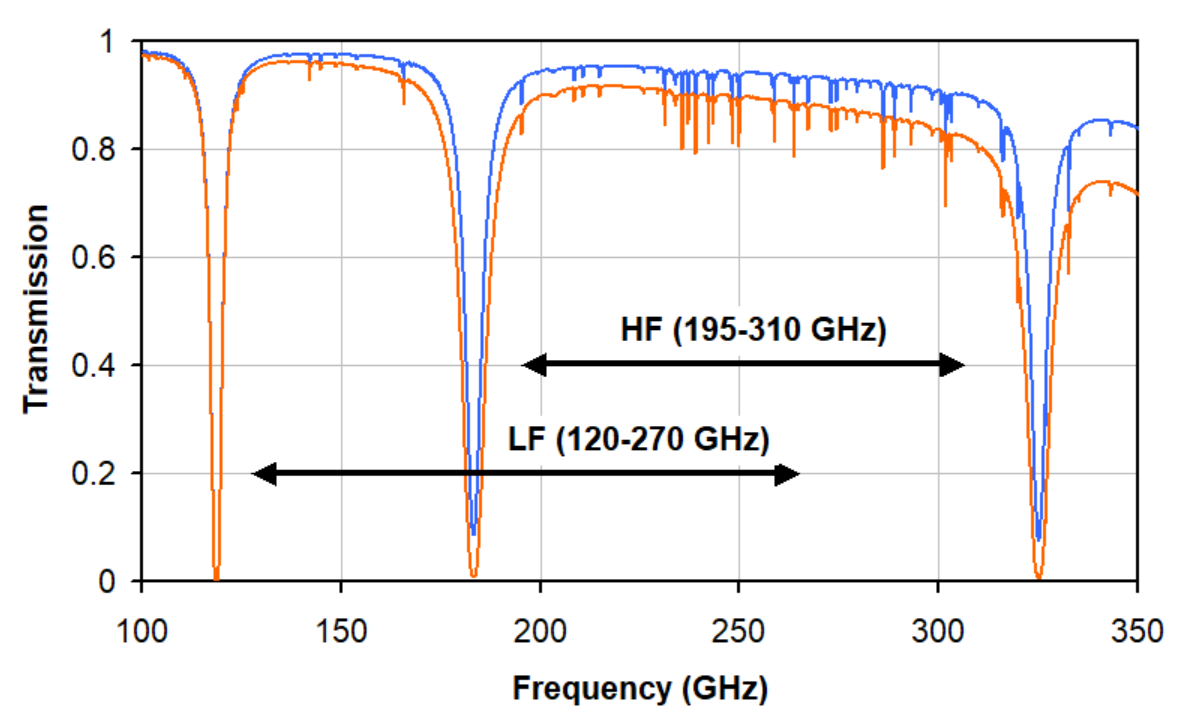}
\end{center}
  \caption{Atmosphere transmission at APEX under 1\,mm PWV (blue) and 2\,mm PWV (orange) conditions. The CONCERTO bands for the HF and LF arrays are shown.}
\label{atmosphere}
\end{figure}

The Atacama Pathfinder EXperiment (APEX) telescope is a modified prototype ALMA antenna with a primary mirror diameter of 12\,meters and a usable field-of-view of about 20\,arcminutes. The location at around 5100\,meters a.s.l. on the Chajnantor plateau ensures optimal observing conditions (see Fig.\,\ref{atmosphere}). In particular, the fraction of time showing a precipitable water vapour (PWV) column lower than 2\,mm is of the order of 70\% or more. 

The structure of the Cassegrain cabin, and in general the telescope infrastructure, had been designed to host large field-of-view instrumentation. The primary mirror surface has been recently refurbished, and achieves in some conditions a precision of the order of 10\,$\mu m$ RMS. APEX is thus, to date, a state-of-the-art installation for millimetre and sub-millimetre Astronomy. The telescope has hosted since 2007, in the same place that will be occupied by CONCERTO, the Large APEX BOlometer CAmera (LABOCA) operating at 360\,GHz \citep{Siringo2009}. 

\begin{figure}[ht]
\begin{center}
\includegraphics[width=9cm,angle=0]{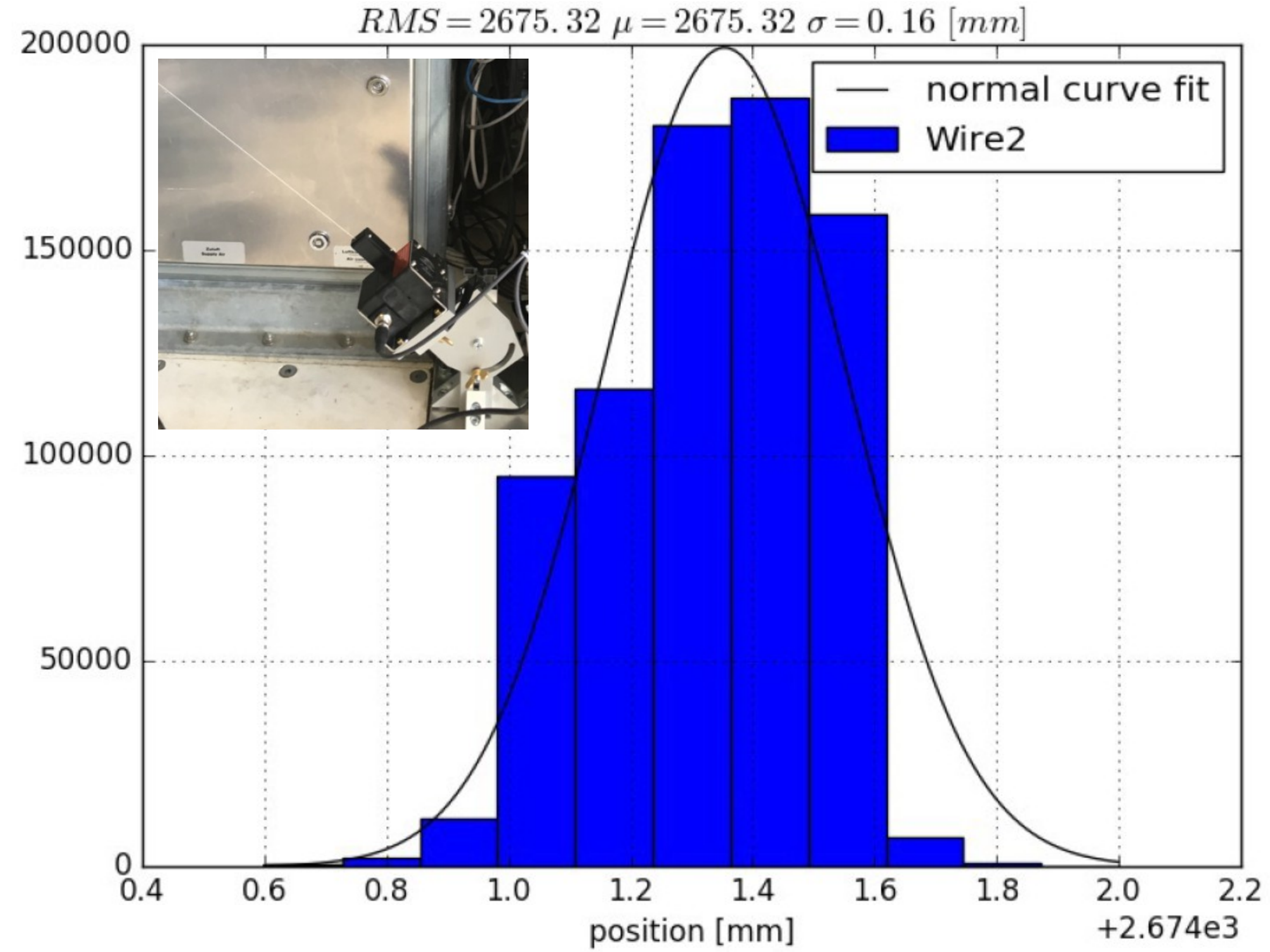}
\end{center}
\caption{Deformations of a wire sensor running between the floor and the top of the APEX Cassegrain cabin recorded during 48\,hours. The average absolute length of the wire is 2675.3\,mm. Inset: picture of the wire sensor.}
\label{deformations}
\end{figure}

The CONCERTO optics box and chassis are slided separately through the C-cabin door and then fixed to the floor by a sufficient number of 12\,mm metric screws. Since the beam will bounce between the floor and the top of the cabin (M4), we have measured the deformations of the C-cabin itself under typical APEX observing conditions. This measurement was achieved using two linear wire sensors with range of 3\,meters and a single measurement precision of 0.1\,mm. As shown in Fig.\,\ref{deformations}, the RMS of both sensors, over 48\,hours of data taken during standard APEX observations, is smaller than 0.2\,mm and thus negligible for our purposes. 

The alignment of the mirrors in the optics box will be achieved in laboratory. The position of each mirror is adjusted with three micro-metric screws. No tuning will be possible at the telescope. The alignment of the optics box with respect to the chassis is ensured by the mechanical fixations. A set of specific lasers will be mounted to achieve in laboratory the internal alignments. At the telescope, we will use these lasers to align the optics box with respect to M4, M4 with respect to M3 and M3 with respect to M2. The alignment procedures will represent a critical step of the installation.

\section{Detectors laboratory characterisation}\label{sec3}

We describe in this section the first tests done on CONCERTO detectors. Some of the electrical tests on the resonances occurred in the CONCERTO cryostat itself. On the other hand, the optical characterisation of the detectors has been achieved in the former NIKA2 test-bench. The so-called NIKA1.5 camera is an easily re-configurable optical dilution cryostat with a base temperature of 60\,mK. It has been recently modified to host one CONCERTO array at a time. In particular, the optical filters can be easily replaced, and NIKA1.5 can be interfaced to a custom MpI for spectral characterisation, or alternatively to a Sky Simulator (described in detail in \citealt{nika1}) for sensitivity and beams geometry measurements. 

\begin{figure}[ht]
\begin{center}
\includegraphics[width=9cm, angle=0]{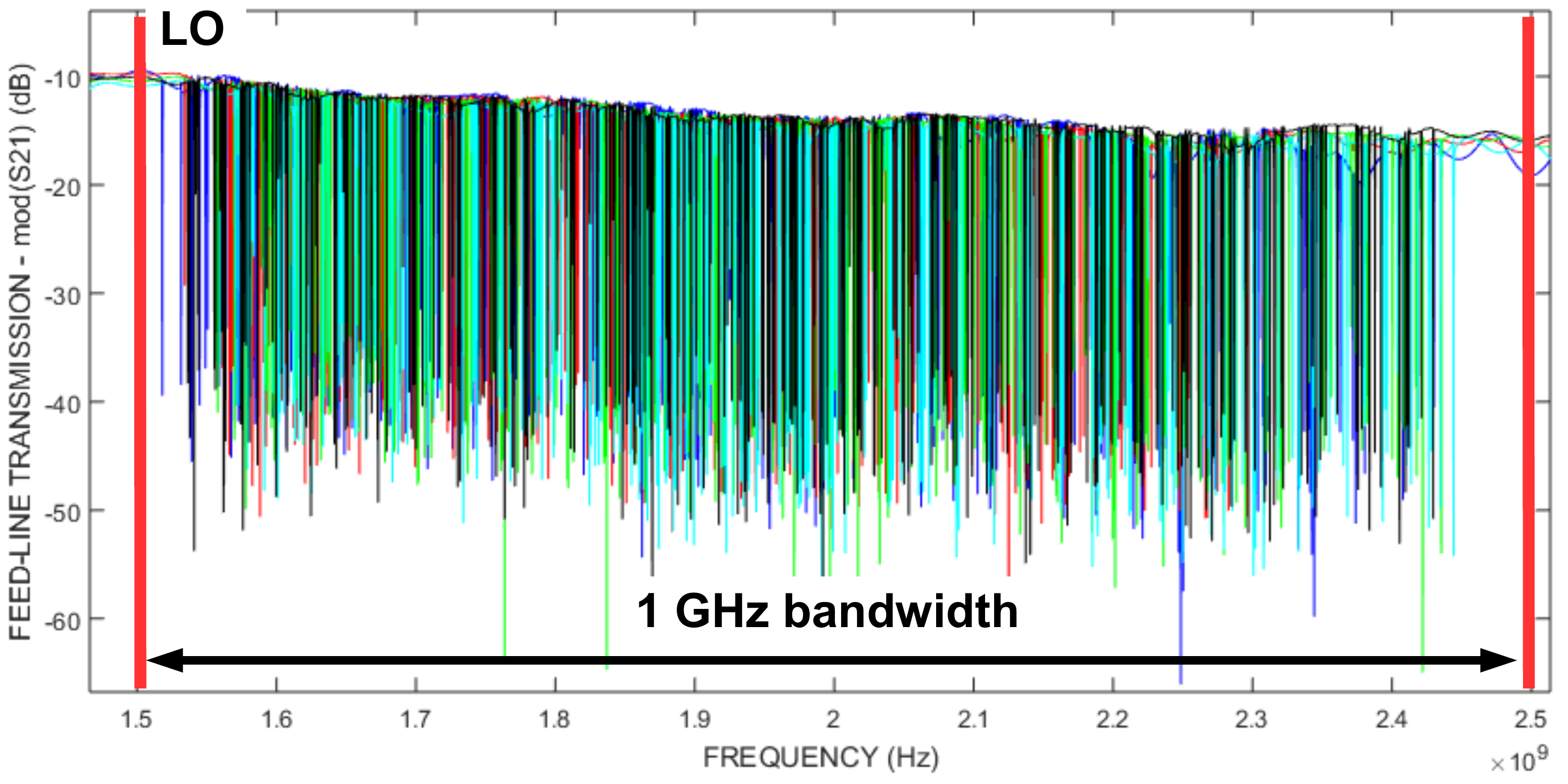}
\end{center}
  \caption{Frequency sweep (transmission of the feed-line between port 1 and 2, i.e. mod(S21)) of five blocks of resonances out of the baseline HF array. They have been acquired under dark conditions, in the CONCERTO cryostat, at T=70\,mK.}
\label{S21}
\end{figure}

Since the twelve readout lines of the HF and LF arrays share a common Local Oscillator (LO, frequency reference for the readout electronics), it is of vital importance to accommodate all the resonances block in a common $\leq$1~GHz band. This is nicely achieved, for example, in the case of the HF array shown in Fig.\,\ref{S21}. The spread between blocks of resonances belonging to the same array, and between different arrays, is mainly due to inhomogeneities and uncertainties in the thickness of the Aluminium film.

\begin{figure}[ht]
\begin{center}
\includegraphics[width=9cm, angle=0]{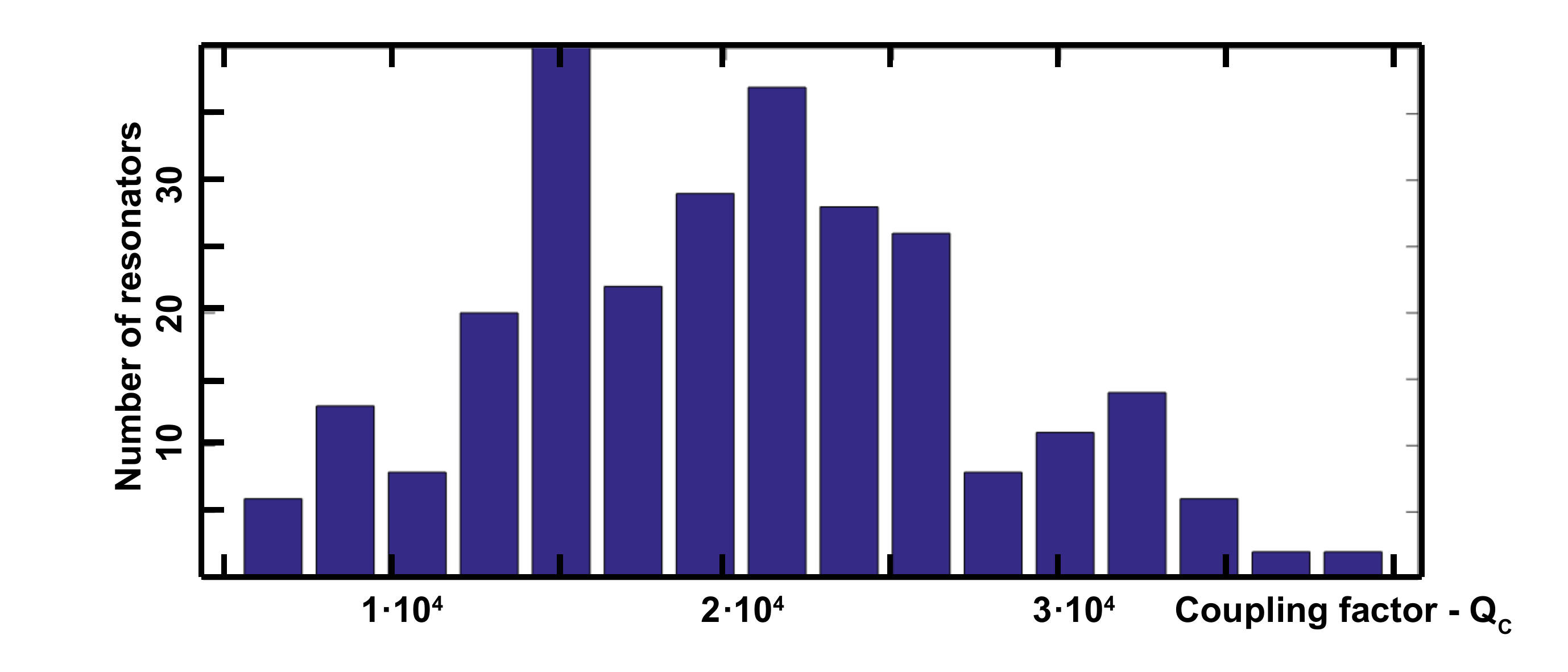}
\end{center}
  \caption{Quality factors distribution for one representative block of resonances in the CONCERTO HF array. For this particular block Q$_c$=23k+/-12k, in line with all the other blocks and with the designed Q$_c$=25k. These quality factors have been measured in the CONCERTO cryostat at T=70\,mK.}
\label{Q}
\end{figure}

Another important electrical parameter to be studied for large arrays of LEKID is the coupling quality factor Q$_c$. The micro-strip configuration that has been chosen has the advantage of guaranteeing a relatively well-peaked distribution of Q$_c$. This is achieved without requiring complicated and risky additional technology steps like cross-the-line micro-bondings or suspended micro-bridges. Figure \,\ref{Q} shows an example of the statistics obtained for one readout line of the CONCERTO HF array. All the lines, as well as the LF arrays tested so far, exhibit similar behaviours. The quality factor distribution is, as designed, peaking around 25k.

\begin{figure}[ht]
\begin{center}
\includegraphics[width=9cm, angle=0]{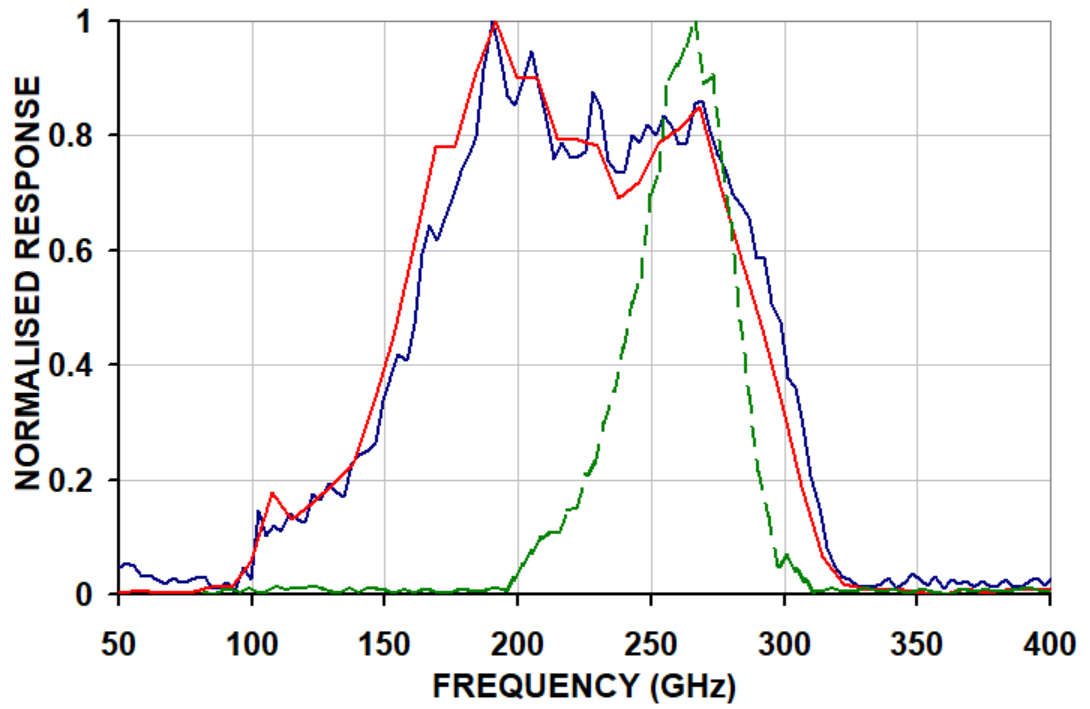}
\end{center}
  \caption{Spectral response of two CONCERTO HF arrays (solid lines) and one NIKA2 260\,GHz detector (dashed line). Red: HR Silicon substrate thickness of 110$\pm$5$\mu$m; blue: thickness of 100$\pm$5$\mu$m.  Spectral responses have been measured in the NIKA1.5 cryostat with low-pass filters defining an open band up to 300\,GHz.}
\label{Spectra}
\end{figure}

The spectral response of two HF arrays, with slightly different substrate thicknesses, has been measured. The results are reported in Fig.~\ref{Spectra}. 

\begin{figure}[ht]
\begin{center}
\includegraphics[width=9.5cm, angle=0]{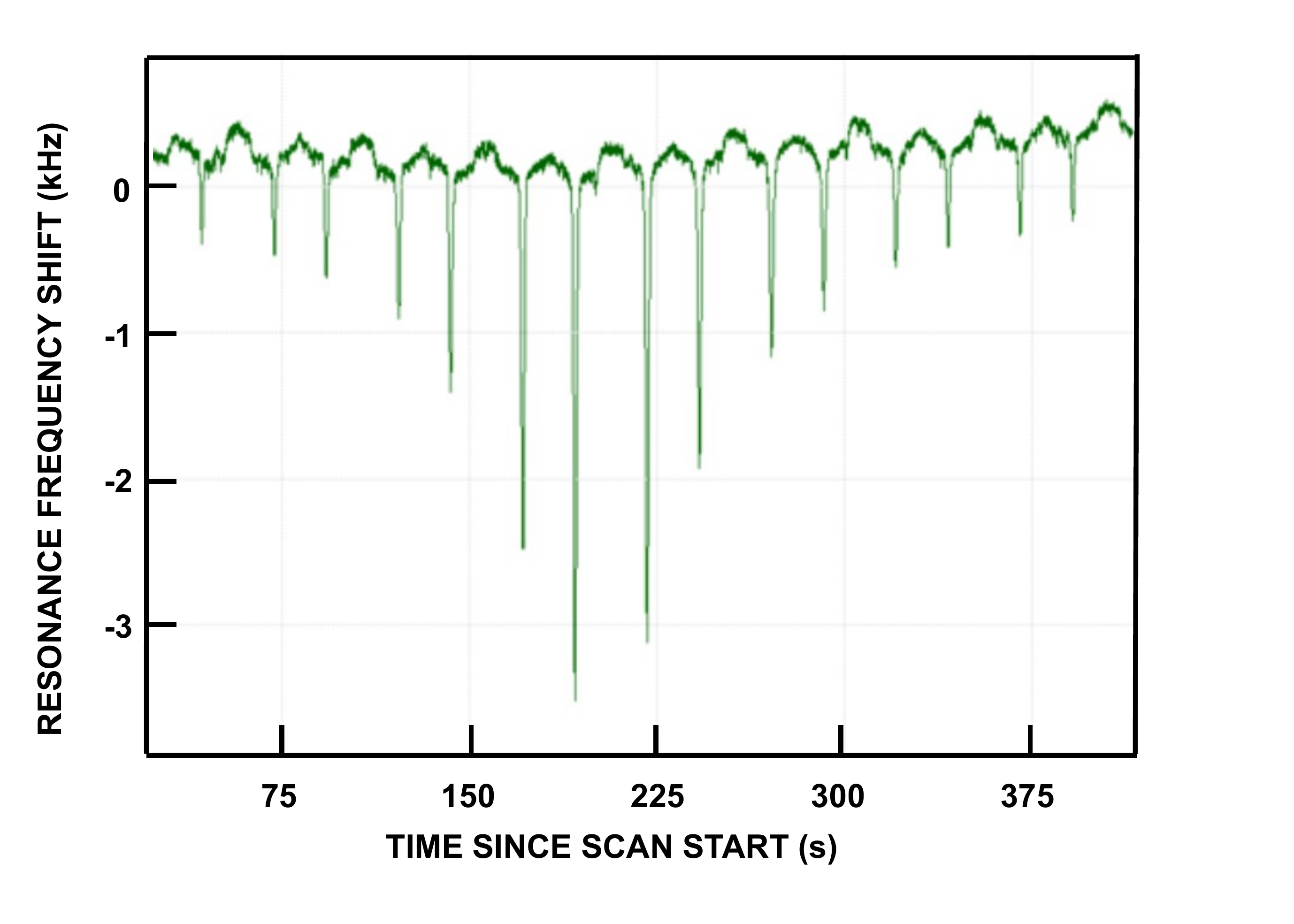}
\end{center}
  \caption{Sky simulator trace. A fake "planet" (point-like source) is crossing the field-of-view of the considered pixel. A raster scan with sub-scans at fixed elevation is simulated. The "elevation" steps are of 4\,mm each. This measurement has been obtained using the NIKA1.5 cryostat at T=120\,mK and under a background temperature around 50\,K.}
\label{Planet}
\end{figure}

The sensitivity has been measured in terms of NET (Noise Equivalent Temperature), specific for the NIKA1.5 optics system. The average NET per pixel of the CONCERTO arrays, in NIKA1.5, is around 2\,mK/$\sqrt{Hz}$. This results in an NET of about 45\,$\mu$K/$\sqrt{Hz}$ per array (polarisation), or 32~$\mu$K/$\sqrt{Hz}$\footnote{32~$\mu$K/$\sqrt{Hz}$ is equivalent to 22.6~$\mu$K$\, \mathrm{sec}^{1/2}$.} when combining both polarisations. Since the sensitivities per pixel are in accordance to what had been measured for the very similar NIKA2 detectors \citep{nika2_1}, we will base our sensitivity estimate in Sect.\,\ref{sec4} on NIKA2 values measured on-sky. We believe actually that the sensitivities measured on the maps on the sky for similar detectors are a more realistic prediction compared to somewhat ideal values estimated in laboratory.

The good imaging characteristics of the CONCERTO arrays are demonstrated by the Sky Simulator tests. An example is shown in Fig.\,\ref{Planet}. A deeper geometrical characterisation of the thousands beams is beyond the scope of this paper, and will be only performed on the final arrays. 

The typical response time of the LEKID used for CONCERTO ranges between 30~$\mu$s and 100\,$\mu$s, depending on the background. Even at the chosen sampling rate of 4\,kHz, a cosmic-ray hit will thus represent a single-point glitch in the CONCERTO raw-time traces. The order of magnitude of expected rate is 0.1\,Hz per pixel.

\section{Sensitivity estimates}\label{sec4}
Due to the similarities between the NIKA2 and CONCERTO detectors, we use the NIKA2 sensitivity measured on sky, and on reduced maps, as a base to estimate the sensitivity for CONCERTO. As already advocated, we think that this approach, coupled to our NET laboratory measurements, provides quite realistic predictions.

\subsection{CONCERTO as an photometer: dual-band sensitivity \label{sec4.1}}
We first compute the sensitivity for CONCERTO as if it was a dual-band imager (LF and HF).
For that, we rely on NIKA2 sensitivity measurements on the IRAM 30-meter telescope.
Average NEFDs for NIKA2 (NEFD$_{NIKA2}$) are equal to 9.8 and 36.1 mJy\,sec$^{1/2}$, at 150 and 260\,GHz, respectively, for pwv=2 and an elevation of 60 degrees \citep{nika2_2}. These numbers already suffer from the transmission of the whole experiment. \\
We observe a large difference between the NIKA2 260 and 150\,GHz channel performances. A combination of known effects explains the gap in sensitivity. Indeed, at 260\,GHz, i) the beam efficiency of the 30-meters telescope is about 55\%, ii) the sensitivity is strongly affected (by 35\%) by a known defect of the NIKA2 dichro\"ic, and iii) the contribution of residual sky noise to the average NEFDs is important. 
Therefore, as a realistic starting point for CONCERTO, we assume a sensitivity for the LF array equals to that of NIKA2 at 150\,GHz. For the HF array, we assume a sensitivity better than that of the NIKA2 260\,GHz channel, thanks to a gain in beam efficiency at APEX (which is at the order of 80\% at the CONCERTO wavelengths), the lack of dichro\"ic in CONCERTO, and better atmospheric conditions. 
The values are thus (for pwv=2 and an elevation of 60 degrees):
\begin{equation}
NEFD_{NIKA2}^{LF} = 10\,[7.5-15] \,\mathrm{mJy\,sec^{1/2} \,and,}
\end{equation}
\begin{equation}
NEFD_{NIKA2}^{HF} = 15\,[10-20] \,\mathrm{mJy\,sec^{1/2}.}
\end{equation}
The numbers in bracket give the uncertainties on our assumption.
As it was the case for NIKA2, we make the hypothesis that no excess noise will appear in CONCERTO at APEX compared to CONCERTO in laboratory. Of course, this cannot be verified until installation. 

Compared to NIKA2, for CONCERTO, we have to scale the sensitivities to match the APEX telescope size, and we add two polarisers in the optical path (P1 and P2, see Fig.\,\ref{MPIreference}). Sensitivity loss is only due to P1, by a factor between $\sqrt{2}$ (if photon noise dominated) and 2. To be conservative, we consider a factor 2.  
Thus, for a single array of CONCERTO, the NEFD becomes:
\begin{equation}
NEFD^{LF, HF} = NEFD_{NIKA2}^{LF, HF} \times 2 \times \left( \frac{27.5}{11} \right)^2
\end{equation}
where 27.5 and 11 meters are the IRAM and APEX telescopes effective sizes, respectively, i.e. the portion of the primary mirrors that are optically conjugated to the cold pupils of the instruments (aperture stops).

Then we assume a frequency window of $\Delta\nu$=115\,GHz, making the assumption that the two arrays cover the frequency range 195-310\,GHz (HF) and 130-270\,GHz (LF), with a notch filter removing 25\,GHz of the low-frequency bandpass (around 183\,GHz). Finally, we also have to take into account the decrease in transmission due to increase optics complexity of CONCERTO compared to NIKA2 (in particular FTS optics will have some transmission loss and additional loading), which we estimate to be $T=0.8$ for an unpolarised source (this is only an additional loss of transmission compared to NIKA2+P1 and not the overall transmission).
\par\medskip\noindent
The sensitivity of CONCERTO as a dual-band photometer (set when the optical path difference in the FTS is null) is thus:
\begin{equation}
NEFD_{\mathrm{phot}}^{LF, HF} = NEFD^{LF, HF} \times \sqrt{\frac{\Delta\nu_{NIKA2}^{LF, HF}}{\Delta\nu}} \times \frac{1}{\sqrt{T}}
\end{equation}
We have $\Delta\nu_{NIKA2}^{HF}$=48\,GHz and  
$\Delta\nu_{NIKA2}^{LH}$=39.2\,GHz \citep{nika2_2} and thus:\\

\noindent $NEFD_{\mathrm{phot}}^{LF}$ = 81.6\, [61.2 - 122.4] \,mJy\,sec$^{1/2}$ and\\ 
$NEFD_{\mathrm{phot}}^{HF}$ = 135.4\, [90.3 - 180.6]\,mJy\,sec$^{1/2}$. \\
 
Note that due to the FTS in front of the cryostat, CONCERTO is a non-optimal instrument for imaging. However, the option of removing the first polariser for purely photometric campaigns could be studied. In that case the sensitivity is expected to be 2 times better for each individual array (as we took a factor 2 of penalty for P1).

\subsection{Sensitivity in spectroscopy}

For the spectroscopic mode, we consider a fix value for spectral resolution $\delta\nu$=1.5\,GHz.
The number of spectral elements in the frequency range is $N_{se}$= $\Delta\nu$/$\delta\nu$.
\par\smallskip\noindent
The sensitivity per spectral element (in mJy s$^{1/2}$) for a single spectrometer (note that in our case, with an FTS, the number of pixel equals the number of spectrometers) is given by
\begin{equation}
NEFD_{FTS}= NEFD_{\mathrm{phot}} \times N_{se} \,.
\label{Eq_1}
\end{equation}

\noindent The beam area is computed assuming a Gaussian beam,
\begin{eqnarray}
\Omega_{\mathrm{beam}} &=& 2\pi \left( \frac{\theta_{\mathrm{beam}}}{2\sqrt{2 \log{2}}} \right)^2\,,
\end{eqnarray}
with a FWHM determined by the Rayleigh criterion for a D=11m antenna (our illumination of the APEX 12m antenna) at a given frequency (corresponding to a given redshift for the [CII] line),
\begin{eqnarray}
\theta_{\mathrm{beam}} &=& 1.22\lambda_{\mathrm{obs}} / D\,.
\end{eqnarray}

\noindent We can then convert the sensitivity per spectral element from point source (Eq.\,\ref{Eq_1}) to diffuse emission (in MJy\,sr$^{-1}$\,s$^{1/2}$) following
\begin{equation}
NEI_{FTS} = NEFD_{FTS} \times 10^{-9} / \Omega_{\mathrm{beam}}
\label{Eq_2}
\end{equation}
This is the Noise Equivalent Intensity, on sky, per KIDS, per spectral bin (taken as $\delta\nu$=1.5\,GHz).

\noindent  We can finally compute the mapping speed MS (per spectral element) following,
\begin{equation}
\mathrm{MS} = \mathrm{FOV} / NEI_{FTS}^2
\end{equation}
where FOV is the field of view area (with a diameter of 20\,arcminutes). Numbers are given in Table\,\ref{Tbl_sens}. Note that we ignored the frequency overlap between the two arrays (and thus a gain of $\sim \sqrt{2}$ on the sensitivity in the frequency overlap region). We considered $NEFD_{DB}^{LF}$ for $\nu \le$ 150\,GHz, $NEFD_{DB}^{HF}$ for $\nu \ge$ 260\,GHz, and a linear interpolation between the two NEFDs for $150<\nu<260$\,GHz.\\

\noindent  We also give in Table\,\ref{Tbl_sens} the sensitivity for the whole array, per spectral element, which is: 
\begin{equation}
\sigma_{\mathrm{array}} = NEI_{FTS} / \sqrt{N_{\mathrm{KIDS}}}\,,
\label{sigma_array}
\end{equation}
where N$_{\mathrm{KIDS}}$ is the number of pixels (KIDS) of each array (we use 1720 KIDS, which correspond to 80\% of valid KIDS in each array). This would be the sensitivity of each voxel\footnote{A voxel represents a value on a regular grid in three-dimensional space.} of large maps, assuming a RA-DEC (or AZ-EL) raster scan-like scanning strategy (and assuming pixel sizes of the map equal to beam sizes). Each voxel of the observed map would then be observed by each KIDS. We checked these numbers using a scanning strategy similar to NIKA2 raster scans, with 3 interferograms per beam. 

\begin{table*}[!h]
\begin{center}
\begin{tabular}{l|cccccc}
$\nu$ [GHz] & 131 &  156 & 211 & 238 & 272 & 302\\ \hline
Redshift of the [CII] line & 13.5 &  11.2 &  8.0 &  7.0 &  6.0 &  5.3 \\ \hline
Beam size [arcsec] & 52.4 & 44.0 & 32.5 & 28.8 & 25.2 & 22.7 \\ \hline
Beam solid angle [$\times$10$^{-8}$ sr] & 7.30 & 5.15 & 2.81 & 2.21 & 1.70 & 1.37 \\ \hline

Mapping Speed & 42.8 [19.0-76.0] & 19.8 [9.0-35.9] & 3.4 [1.8-7.1] & 1.7 [0.9-3.7] & 0.8 [0.5-1.9] &  0.5 [0.3-1.2] \\ 

[$\times$10$^{-3}$ deg$^2$ / (MJy/sr)$^2$ / hour]   & & & &  & &  \\ \hline

On sky map sensitivity $\sigma_{array}$ & 2.1 [1.5-3.1] & 3.0 [2.3-4.5] & 7.3 [5.1-10.2] &  10.4 [7.1-14.1] & 14.8 [9.9-19.7] & 18.2 [12.2-24.3] \\ 

[(MJy/sr) $\mathrm{sec}^{1/2}$] & & & & & & \\

\end{tabular}
\caption{\label{Tbl_sens} Key parameters for CONCERTO instrument. Sensitivities and mapping speeds are given for one spectral element (with $\delta\nu$=1.5\,GHz), assuming 80\% of valid KIDS, a precipitable water vapour of 2 mm and an elevation of 60 degrees.}
\end{center}
\end{table*}

\section{Low spectral-resolution spectroscopic surveys}\label{sec5}

CONCERTO will offer a generic access to a large FoV and low-frequency resolution spectroscopic instrument. This opening of 3D large-scale surveys is the next step after the broad-band photometric experiments, either from the ground (e.g. LABOCA, SCUBA2, NIKA2) or from space (e.g. Herschel and Planck). 
The first scientific aim of CONCERTO is to map in three dimensions the fluctuations of the [CII] line intensity in the reionisation and post-reionisation epoch (z$\ge$5.3). This technique, known as "intensity mapping", will measure the clustering of [CII] emissivity and allow answering questions on how and when galaxies and quasars formed, and the history and topology of reionisation. Even if [CII] intensity mapping has been the basis of instrument definition, we extended the instrument capabilities to make CONCERTO a multi-purpose instrument (e.g., extending the frequency range down to 130\,GHz for galaxies clusters observations). Thus we expect CONCERTO to bring a significant contribution in a number of areas, including the study of galaxy clusters (via the thermal and kinetic SZ effect), the follow-up of  cosmological deep surveys, the observation of local and intermediate-redshift galaxies, and the study of Galactic star-forming clouds. In this section we give some forecast on the expected signal to noise ratio that can be obtained on the [CII]-emission power spectrum (Sect.\,\ref{CII_IM}). In addition, we give some predictions for observing the SZ signal of galaxy clusters (Sect.\,\ref{SZ}).

\subsection{\label{CII_IM} [CII] intensity mapping with CONCERTO}
[CII] is one of the brightest emission lines in the spectra of galaxies. It is an excellent coolant for neutral gas in photo-dominated regions and an extinction-free tracer of star formation at high $z$.
Being redshifted into the sub-millimetre and millimetre atmospheric windows for $z>4.5$, it has become one of the most popular line at high $z$. Pointing on known objects, with e.g. ALMA, NOEMA, APEX/FLASH, [CII] is now detected in a large number of galaxies at high $z$ ($>$150 at $z>4.5$, with a large contribution from the ALMA ALPINE survey, e.g. \citealt{bethermin2020}). Such observations are a tremendous step forward but we also need to look at the overall population, i.e. observe large volume and unbiased surveys. First observations with ground-based interferometers e.g., with ALMA ("ALMA Spectroscopic Survey in the Hubble Ultra Deep Field" large program - ASPECS, \citealt{walter2016}) or JVLA (CO Luminosity Density at High Redshift  survey - COLDz, \citealt{riechers2019}) offer a three-dimensional view of the molecular gas content of galaxies. The covered areas are about 5-60 square arcminutes. \\

Intensity mapping complements these efforts beautifully \citep[e.g][]{kovetz2019} providing an unbiased view of the distribution of the gas that is difficult to assemble from targeted measurements of individual galaxies, and probing cosmological volumes, with maps on several-degree scale and large frequency (and thus redshift) coverage. Intensity mapping exploits the confusion-limited regime and measures the integrated light emission from all sources, including unresolved faint galaxies. \\

We will conduct with CONCERTO a major survey of about one square degree with 1,200\,hours of APEX telescope time.
The survey will provide a data cube in which intensity is mapped as a function of sky position and redshift. 
Our main target is the [CII] line emission at $z\ge5.3$. But CONCERTO will also observe the CO intensity fluctuations arising from $0.3<z<2$ galaxies, giving the spatial distribution and abundance of molecular gas over a broad range of cosmic time. 
The 3-D fluctuations will be studied in Fourier space with the power spectrum.\\

To compute the expected SNR on the [CII] power spectrum at high $z$, we used the [CII] model presented in \cite{Serra2016}.
Using measurements of the Cosmic Infrared Background (CIB) angular power spectra from Herschel/SPIRE together with star formation rate density (SFRD) measurements, they constrain the galaxy FIR luminosity as a function of dark-matter halo mass at all relevant redshifts, in the halo model framework. By using scaling relations from \cite{Spinoglio2012} to link the intensity of emission lines to the galaxy infrared luminosity, they compute 3D emission line power spectra for all relevant lines, including [CII]. They compute the expected SNR of cross-power spectra between [CII] and other emission lines, that will constrain the mean amplitude of each signal, and thereby gain insight into the mean properties of the ISM of high-$z$ galaxies. Note that in their paper, they use for CONCERTO a constant sensitivity for all redshifts of $\sigma_{array}$= 155\,mJy\,sec$^{1/2}$, while we have here $\sigma_{array}$=[156, 206, 230,  250,  250] mJy s$^{1/2}$ at z=[11.2,  8.0, 7.0, 6.0, 5.3].\\

We follow \cite{gong2012} to compute uncertainties on the power spectra.
The observing time per map voxel (considering one pixel equals one beam) is given by
\begin{eqnarray}
t_{\mathrm{voxel}} &=& t_{\mathrm{survey}} \frac{\Omega_{\mathrm{beam}} N_{\mathrm{KIDS}}}{A}\,,
\end{eqnarray}
with A the survey area, $t_{\mathrm{survey}}$ the on-sky survey time (i.e. 1200$\times$0.7=840\,hours in our case, considering 30\% of overheads), $\Omega_{\mathrm{beam}}$ the solid angle of the beam (Table\,\ref{Tbl_sens}) and $N_{\mathrm{KIDS}}$, the number of pixels (we consider 80\% of valid KIDS, thus $N_{\mathrm{KIDS}}$=1720).

\noindent Assuming a spherically averaged power spectrum measurement, and
a directionally independent on sky sensitivity, 
the variance of the power spectrum is:
\begin{eqnarray}
\mathrm{var}[\bar{P}_{CII}(k)] = \frac{[P_{CII}(k)+
\bar{P}_{CII}^{\mathrm{N}}(k)]^2}{N_{\mathrm{m}}(k,z)},
\label{eqn:avg_spectrum}
\end{eqnarray}
where  $\mathrm{N_m(k,z)}$ is the number of modes that leads to the power spectrum measurement at each $k$ and
\begin{eqnarray}
\bar{P}_{CII}^{\mathrm{N}}(k) = V_{\mathrm{voxel}} \frac{\sigma_{voxel}^2}{t_{\mathrm{voxel}}}\,,
\label{PCII}
\end{eqnarray}
with $V_{\mathrm{voxel}}$ the volume surveyed by each voxel. In the case of CONCERTO, each KIDS gets a spectrum, and considering one KID per beam, we have $\sigma_{voxel}$ = $\sigma_{array}$. 

The number of modes at each $k$ is given by
\begin{eqnarray}
N_m(k,z) = 2\pi\,k^2\Delta\,k\frac{V_s(z)}{(2\pi)^3}\,,
\end{eqnarray}
where $\mathrm{\Delta\,k}$ is the Fourier bin size, 
and V$_\mathrm{s}$(z) the survey volume, expressed as
\begin{eqnarray}
V_s(z) = \chi(z)^2 y_{CII} B_{\nu} A\,,
\label{Vs}
\end{eqnarray}
with
\begin{eqnarray}
y_{CII}(z) &=& \lambda_{CII}(1+z)^2/H(z),
\end{eqnarray}
being the factor to convert the frequency intervals to the comoving distance at the wavelength $\lambda_{CII}$ (rest frame wavelength of the [CII] line). In Eq.\,\ref{Vs}, $B_{\nu}$ is the considered bandwidth of the measurement. The volume surveyed by each voxel ($V_{\mathrm{voxel}}$ in Eq.\,\ref{PCII}) is
\begin{eqnarray}
V_{\mathrm{voxel}}&=&\chi(z)^2 y_{CII}(z)\Omega_{\mathrm{beam}}\delta_\nu \,.
\end{eqnarray}

\noindent We consider measurements spanning a redshift range $\mathrm{\Delta\,z\sim0.6}$
which corresponds to a frequency range of $\mathrm{B_{\nu}}\sim20$ GHz at  $z=6.1$ for the [CII] line. Note that \cite{gong2012} give useful relations for computing $V_s$ and $V_{\mathrm{voxel}}$ for the [CII] line.\\

\begin{figure}[ht]
   \centering
  \includegraphics[width=0.5\textwidth]{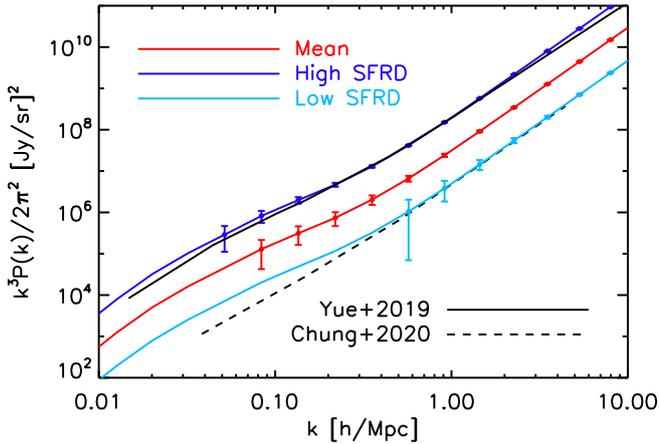}
   \caption{Predicted [CII] power spectrum at z=6. Three cases are shown, corresponding to three scenarios of SFRD at high $z$ \citep[see][]{Lagache2018_proc}: High SFRD (dark blue), Low SFRD (light blue) and the geometrical mean of the two (red). Only points with SNR$>$1 are shown. SNRs have been computed considering A=1.4 square degrees, $t_{survey}$=840 hours (which corresponds to a total observation time of 1200 hours taking into account the overheads) and sensitivities as estimated in Sect.\,\ref{sec4}. [CII] power spectra have been derived from the modelling of CIB power spectra \citep{Serra2016}, using a conversion from SFR to [CII] that conservatively underestimates the [CII] luminosity by a factor 6 at z=5 compared to recent semi-analytical models \citep[e.g.,][]{Lagache2018} or ALMA ALPINE measurements \citep{Schaerer2020}. Our estimates of [CII] power spectra are thus likely to be underestimated. Also shown are the predicted [CII] power spectra from \citealt{yue2019} (using the local SFR-[CII] relation, black line) and \citealt{chung2020} (dashed black line).}
   \label{fig_pk}
\end{figure}

Table\,\ref{SNR_PK_CII} gives the numbers derived from the above computations and Fig.\,\ref{fig_pk} shows the predicted [CII] power spectrum with its error bars.  Because such predictions are very uncertain, we assumed two extreme models, giving respectively low and high SFRD at z$>$3 (see Fig.\,3 of \citet{Lagache2018_proc}). Low SFRD is the pessimistic prediction as it corresponds to the lowest UV-driven SFRD; high SFRD is the optimistic prediction as it corresponds to the CIB-driven SFRD derived from the halo modelling of Planck CIB measurements \citep{PlanckXXX}. \\

\begin{table*}[ht]
\begin{center}
\begin{tabular}{l|cccc}
Redshift $z$ & 5.5 & 6.2 & 7 & 8 \\ \hline
SNR P$^{CII}$ mean SFRD &  23 [14-44]   &  12 [6.9-24]   &  5.7 [3.1-12]   &   2.0 [1.0-4.1] \\
SNR P$^{CII}$ low SFRD &  4.5 [2.6-9.8]   &  1.9 [1.1-4.3]   &  0.78 [0.48-1.7]   &   0.23 [0.12-0.48] \\
SNR P$^{CII}$ high SFRD &  79 [57-112]  &  55 [36-87]  &  34 [21-60]  &   16 [8.5-29]  \\\hline
\end{tabular}
\caption{\label{SNR_PK_CII} SNR on the [CII] power spectrum, P$^{CII}$, computed for $\Delta$z=0.6 and given for k=[0.1,1] h/Mpc. See the caption of Fig.\,\ref{fig_pk} for more details. Numbers in brackets reflect the range of sensitivities (as given in Table\,\ref{Tbl_sens}).}
\end{center}
\end{table*}

\begin{figure*}[ht]
\includegraphics[width=19cm,angle=0]{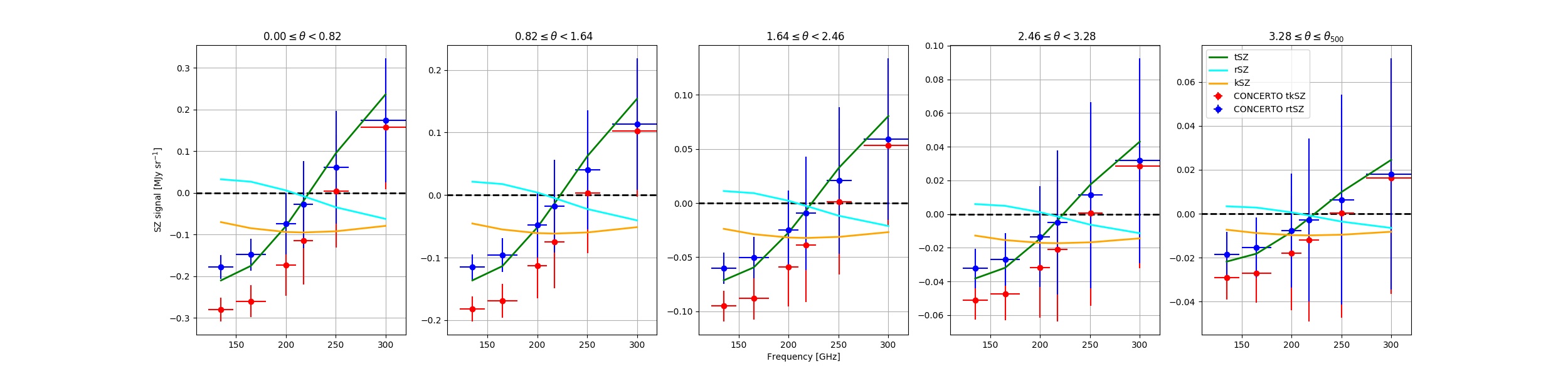}
  \caption{Thermal (green), kinetic (orange) and relativistic (light blue) SZ effect for a simulated cluster at redshift z=0.4, with mass 10$^{15}$\,M$_{\sun}$ and temperature 20\,keV, which is moving at 1000\,km/h towards the observer. From left to right, we present the measured CONCERTO SEDs for tSZ+kSZ and rSZ+tSZ including uncertainties (red and blue dots, respectively) for different radial bins with respect to the center of the cluster ($\theta$ in arcmin) up to the cluster characteristic radius ($\theta_{500}$, which is the radial angular distance at which the mean cluster density is 500 times the critical density at the cluster redshift).}
\label{concerto_sz}
\end{figure*}

Covering an area of 1.4 square degrees, our survey will provide the first measurements of the [CII] power spectrum up to z$\sim$7, considering the mean SFRD and average sensitivity estimate, and up to z$\sim$8 in the best case of sensitivity estimate. Note that the low SFRD case is unlikely as it gives low shot-noise levels for the CIB, not compatible with the Planck and Herschel measurements, and not compatible with current SFRD measurements at high-z based on [CII] or far-infrared measurements. Moreover, on top of the exact level of the SFRD at high $z$, the relation used to convert SFR to [CII] luminosity is another capital ingredient of such model. \cite{Serra2016} used the relation of \cite{Spinoglio2012}, which gives, for a given SFR, a [CII] luminosity 6 times lower than that obtained with the [CII]-SFR relations of \cite{Lagache2018} or observed with the ALMA ALPINE survey \citep{Schaerer2020}. Our power spectra are thus very likely to be underestimated. In terms of point source sensitivities, our survey will reach 1$\sigma$=[14.4, 14.9, 19.7, 22.0, 23.9, 23.9]\,mJy, for a spectral element $\delta \nu$=1.5\,GHz, at $\nu$=[131.0, 156.0, 211.0, 238.0, 272.0, 302]\,GHz, respectively.\\
Finally, one major difficulty in intensity mapping surveys is the problem of foregrounds. For [CII], the main foregrounds will be the contamination from emission lines from lower redshifts, in particular emission from CO rotation transitions \citep[e.g.][] {Yue2015}. \cite{silva2015} and \cite{breysse2015} showed that this contamination can be partially removed by masking out the brightest pixels in the survey, or the low-redshift galaxies selected from other surveys. For that, CONCERTO will highly benefit from the extensive visible-IR photometry and spectroscopy galaxy survey data that are available in the chosen field (i.e. the COSMOS field). In addition, one of the strengths of CONCERTO is in its wide frequency range: several CO lines are simultaneously observed at the same redshift for all redshifts $z>0.35$. Cross-correlation between these lines will be a very powerful method to remove the contamination. This will be specifically addressed for CONCERTO in a future paper.

\subsection{\label{SZ} Observing galaxy clusters with CONCERTO}

Cluster of galaxies are the largest gravitationally bound objects in the Universe and so they are key for understanding the  hierarchical large scale structure \citep{Kravtsov2012}.
The study of galaxy clusters and in particular of their number as a function of mass and redshift allows us to constrain cosmological parameters \citep{Allen2011}. In the frequency range covered by CONCERTO, cluster of galaxies will be mainly detected via the thermal and kinetic Sunyaev-Zel'dovitch (SZ) effect \citep{Sunyaev1972,Sunyaev1980}. The thermal SZ (tSZ) effect \citep{Sunyaev1972} refers to the interaction of the hot electrons in clusters with the CMB photons, and results in a distortion in the CMB spectrum at the position of the cluster. In the case of hot clusters, it will be also affected by relativistic corrections named rSZ \citep[see][for details in their computation]{2004A&A...417..827I}. The kinetic SZ (kSZ) \citep{Sunyaev1980} is a Doppler shift of the CMB photons induced by the proper motion of clusters of galaxies along the line-of-sight. \\

Observations of the tSZ effect have been successfully performed at high angular resolution using continuum cameras based on KIDs, such as NIKA and NIKA2 at the IRAM 30-m telescope \citep{Adam2014, Adam2015,Adam2016a,Adam2017,Ruppin2016,Ruppin2018}. For the kSZ effect, also observed with e.g. NIKA2 \citep{Adam2016b}, a multi-wavelength spectrometer as CONCERTO would be an unique tool to separate the tSZ and kSZ as well as the different foreground components (CIB, CMB and the Galactic emission) and extract information about the cluster physics. Indeed, with sufficiently precise spectroscopy measurements \citep{Birkinshaw1999}, we can measure the cluster mass (from the tSZ effect), proper motion along the line-of-sight (from the kSZ), and temperature (from the relativistic corrections to the tSZ).  \\

Particularly, considering the angular resolution and the mapping speed of CONCERTO, we expect to perform a precise estimate of the shape of the SZ spectrum for clusters of galaxies for redshifts between 0.2 and 0.8. 
As an illustration we present in Fig.\,\ref{concerto_sz} a simulation of a typical cluster that could be observed with CONCERTO. In this simulation the mass of the cluster is equal to 10$^{15}$\,M$_{\sun}$ and the cluster is located at a redshift z = 0.4. 
We assume an universal pressure profile model from \cite{2010A&A...517A..92A} to compute the cluster Compton parameter map. 
From left to right we present the cluster SED as expected to be measured by CONCERTO at different radial distances from the center of the cluster. We have considered 6 bands in frequency with typical bandwidths of 10 to 25\,GHz. We explored both the sensitivity to the relativistic SZ effect (blue points) assuming a cluster temperature of 20 \,keV, and to the kinetic SZ (red points) assuming the cluster is moving towards the observer with a velocity of 1000\,km/h.
Uncertainties were computed from the sensitivity estimates given in Sect.~\ref{sec4} assuming a mapping area of 310\,arcmin$^2$ and a total integration time of 30 hours. We also show in the Fig.\,\ref{concerto_sz} the individual tSZ, kSZ and rSZ effect contributions.  \\
We find that for reasonable observation times (tens of hours) CONCERTO would provide first spectral 2D mapping of the intracluster medium of high redshift clusters and should be able to measure cluster velocities via the kSZ effect. CONCERTO should also be able to detect the relativistic SZ effect and measure cluster temperature. A more detailed mapping of the cluster temperature would require observation times of about hundreds of hours.

\section{Conclusions}\label{sec6} 
We have presented the design of the CONCERTO instrument, a novel spectrometer to be installed on the APEX telescope. CONCERTO is based on the development of new arrays in the millimetre using Kinetic Inductance Detectors. It will contain two arrays of 2152 KIDS, mounted in a dilution cryostat that has a base temperature of 0.1\,K. Spectra are obtained by a fast Martin-Puplett interferometer located in front of the cryostat. Frequency resolution can be up to $\delta_{\nu}$=1\,GHz. 
The technological choices leading to the final instrument design have been explained in detail. Promising detectors characterisation have been obtained. Estimates of expected sensitivity are given, based mostly on the NIKA2 experience, i.e. an instrument on sky subject to similar constraints as CONCERTO. The expected sensitivity, combined with the large field-of-view (20\,arcminute diameter) will provide an unprecedented mapping speed for such an instrument.

CONCERTO will cover the frequency range 130-310\,GHz, allowing in particular the observation of the [CII]-emission line at high redshift and the SZ signal from galaxy clusters. 
These two main science drivers for CONCERTO have been detailed in the paper. CONCERTO is particularly well suited for these two objectives since i) the [CII] line intensity mapping requires a combination of sensitivity, large field-of-view and spectral resolution R$\geq$100 and ii) the study of galaxies clusters via the SZ effect, requires a multi-frequency analysis (to separate the different SZ signals, as well as the SZ signals from CIB, CMB and Galactic dust), the angular resolution of the 12-meters telescope and R$\approx$10.\\
 
CONCERTO is at present in advanced stage of fabrication. The installation and technical commissioning at the APEX telescope is scheduled for the first semester of 2021. The commissioning, science verification and observations are foreseen by the end of 2022.

\begin{acknowledgement}
Besides the authors, the technicians and engineers more involved in the experimental setup development have been Maurice Grollier, Olivier Exshaw, Anne Gerardin, Gilles Pont, Guillaume Donnier-Valentin, Philippe Jeantet, Mathilde Heigeas, Christophe Vescovi, and Marc Marton. We acknowledge the crucial contributions of the whole Cryogenics and Electronics groups at Institut N\'eel and LPSC. The arrays described in this paper have been produced at the PTA Grenoble microfabrication facility. We warmly thank the support from the APEX staff for their help in CONCERTO pre-installations and design. The flexible pipes, in particular, have been routed under the competent coordination of Jorge Santana and Marcelo Navarro.
We acknowledge support from the European Research Council (ERC) under the European Union's Horizon 2020 research and innovation programme (project CONCERTO, grant agreement No 788212) and from the Excellence Initiative of Aix-Marseille University-A*Midex, a French "Investissements d'Avenir" programme. GL warmly thanks Matt Bradford, Jamie Bock and Tzu-Ching Chang for insightful discussions on CONCERTO sensitivity computation and J.-G. Cuby for his help and support for the ERC proposal. We are grateful to our administrative staff in Grenoble and Marseille, in particular Patricia Poirier, Mathilde Berard, Lilia Todorov and Val\'erie Favre, and the Protisvalor team. We acknowledge the crucial help of the Institut Néel and MCBT Heads (Etienne Bustarret, Klaus Hasselbach, Thierry Fournier, Laurence Magaud) during the COVID-19 restriction period. 
\end{acknowledgement}

\bibliographystyle{aa}
\bibliography{conc}

\end{document}